\documentclass[prl,twocolumn,amsmath,amssymb]{revtex4-1}

\usepackage{graphicx}

\usepackage{bm}
\usepackage{color}
\usepackage {cancel}

\begin{document}

\title{Breaking the trade-off between fast control and long lifetime \\of a superconducting qubit}

\author{S. Kono$^{1}$}
\author{K. Koshino$^{2}$}
\author{D. Lachance-Quirion$^{3,4}$}
\author{A. F. van Loo$^{1}$}
\author{Y. Tabuchi$^{3}$}
\author{A. Noguchi$^{5,6}$}
\author{Y. Nakamura$^{1,3}$}%

\affiliation{$^{1}$Center for Emergent Matter Science (CEMS), RIKEN, Wako, Saitama 351-0198, Japan}
\affiliation{$^{2}$College of Liberal Arts and Sciences, Tokyo Medical and Dental University, Ichikawa, Chiba 272-0827, Japan}
\affiliation{$^{3}$Research Center for Advanced Science and Technology (RCAST), The University of Tokyo, Meguro-ku, Tokyo 153-8904, Japan}
\affiliation{$^{4}$Institut quantique and Département de physique, Université de Sherbrooke, Sherbrooke, Québec, J1K 2R1, Canada}
\affiliation{$^{5}$Komaba Institute for Science (KIS), The University of Tokyo, Meguro-ku, Tokyo, 153-8902, Japan}
\affiliation{$^{6}$PRESTO, Japan Science and Technology Agency, Kawaguchi-shi, Saitama 332-0012, Japan}

\date{\today}

\begin{abstract}
The rapid development in designs and fabrication techniques of superconducting qubits has made coherence times of qubits longer.
In the near future, however, the radiative decay of a qubit into its control line will be a fundamental limitation, imposing a trade-off between fast control and long lifetime of the qubit.
In this work, we successfully break this trade-off by strongly coupling another superconducting qubit along the control line.
This second qubit, which we call a Josephson quantum filter~(JQF), prevents the qubit from emitting microwave photons and thus suppresses its relaxation, while faithfully transmitting large-amplitude control microwave pulses due to the saturation of the quantum filter, enabling fast qubit control.
We observe an improvement of the qubit relaxation time without a reduction of the Rabi frequency.
This device could potentially help in the realization of a large-scale superconducting quantum information processor in terms of the heating of the qubit environment and the crosstalk between qubits.
\end{abstract}

\maketitle

Single-qubit gates are an essential element for any quantum protocol based on qubits~\cite{Nielsen2010}. 
They are typically implemented by applying an electromagnetic field in resonance with the energy difference between two levels, inducing Rabi oscillations~\cite{Monroe2013,Awschalom2013,Devoret2013}.
The qubit has to be coupled to at least one one-dimensional continuous mode to have an external control. 
Although a larger coupling strength to the control degree of freedom achieves a faster gate operation for a given drive amplitude, it also increases the radiative decay of the qubit into that continuous mode.
Conversely, suppressing this radiative decay by reducing the coupling strength leads to slower qubit control.
This is a fundamental trade-off between fast control and long lifetime of a qubit, which originates from the fluctuation-dissipation theorem~\cite{Gardiner2004}.

Superconducting qubits are one of the promising candidates for a large-scale quantum processor~\cite{Otterbach2017,Ye2019,Song2019,Arute2019}.
The ceaseless developments in designs and fabrication techniques have been extending coherence times of the qubits~\cite{Kjaergaard2019,Paik2011,Pop2014,Yan2016,Nguyen2019,Gyenis2019}.
The radiative decay of a superconducting qubit to its control line can no longer be dismissed in devices with state-of-the-art coherence times.
The trade-off in qubit control has so far been dealt with by designing a weak coupling to the control line and applying a strong microwave drive field for compensation~\cite{Krinner2019}. 
However, further improvements in the coherence time of superconducting qubits would require even weaker coupling to the control line, leading to an increase in the microwave power needed to control the qubits.
This will be problematic for large-scale superconducting quantum circuits due to heating of the qubit cryogenic environment~\cite{Krinner2019,Yeh2017,Ikonen2017,Wang2019} and the output power level of the control electronics~\cite{Krantz2019,McDermott2018}.
Furthermore, the demand for a strong microwave drive field may increase crosstalk to non-targeted qubits in the vicinity~\cite{Sheldon2016}.

Here, we experimentally demonstrate the suppression of the radiative decay of a ``data'' qubit to its control line without sacrificing the gate speed by using an ancillary qubit that acts as a nonlinear filter.
We name this new type of qubit filter a Josephson quantum filter~(JQF)~\cite{Koshino2020}.
As shown in Fig.~1(a), on one hand, the JQF reflects single photons emitted from the data qubit~\cite{Astafiev2010,Hoi2011}, suppressing the radiative decay to the control line.
On the other hand, when a large-amplitude control microwave field is applied~[Fig.~1(b)], the JQF saturates and becomes transparent~\cite{Astafiev2010,Hoi2011}, enabling fast Rabi oscillations of the data qubit.
The working principle is in stark contrast to that of a Purcell filter, which utilizes the frequency difference between a qubit and a readout resonator to realize both fast readout and long lifetime of the qubit~\cite{Houck2008,Reed2010APL,Jeffrey2014,Sete2015}.The Purcell filter circuit is not suitable, however, for a case where the frequencies of the radiative decay and the control signal are identical.

\begin{figure*}[!t]
\begin{center}
  \includegraphics[width=130mm]{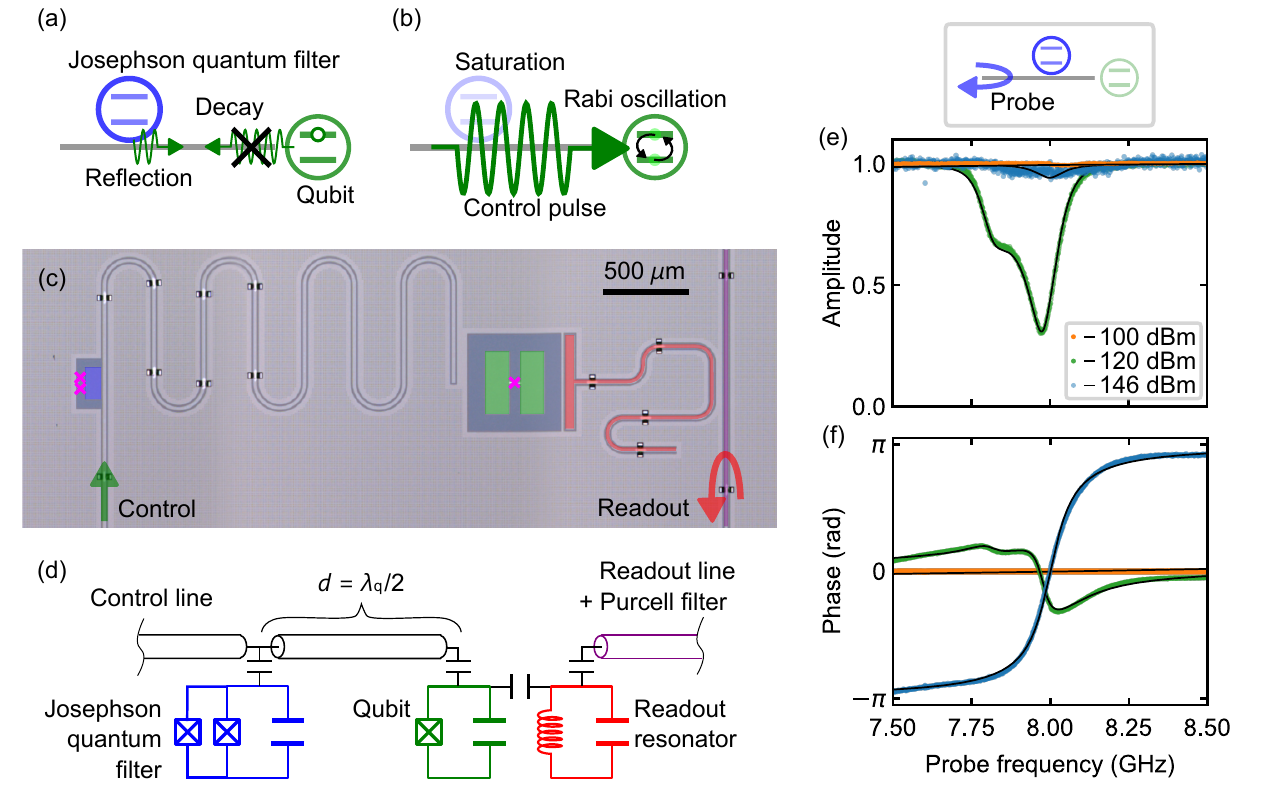} 
\caption{Josephson quantum filter~(JQF).
(a), (b)~Concept of the JQF.
The JQF reflects single photons emitted from the qubit, suppressing the qubit radiative decay.
On the other hand, when a large-amplitude control pulse is applied to the qubit, the JQF becomes saturated and transmits the pulse, enabling fast qubit control.
(c)~False-colored optical image and (d)~equivalent circuit of the fabricated superconducting circuit.
A fixed-frequency transmon qubit~(green) acting as the data qubit is connected to a control coplanar waveguide with an ancillary tunable-frequency transmon qubit~(blue) with a SQUID acting as the JQF.
The data qubit is coupled to a resonator~(red) for fast dispersive readout via a readout line with a Purcell filter~(purple).
Air-bridges are fabricated on the waveguides to suppress spurious modes.
The Josephson junctions are indicated by magenta crosses.
(e)~Amplitude and (f)~phase of the reflection spectra of the JQF measured via the control line.
The dots~(orange, green, cyan) are the experimental results with different probe powers, and the black lines are the theoretical fits.
The probe power of $-120$~dBm corresponds to the single-photon power level for the JQF, defined as $\hbar\omega_{\rm f}(\gamma_{\rm ex}^{\rm f}+\gamma_{\rm in}^{\rm f})^2/4\gamma_{\rm ex}^{\rm f}$, which would populate a linear resonator with a single photon on average.
Note that the qubit transition is not observed here since the resolution of the probe frequency is larger than the qubit linewidth.
} 
  \label{fig1} 
\end{center}
\end{figure*}

A system composed of a qubit and a JQF in a semi-infinite control line is described theoretically by the waveguide-quantum-electrodynamics formalism~\cite{Roy2017,Lehmberg1970,Chang2012,Lalumiere2013,van2013,Mirhosseini2019,Kannan2019}.
As shown in Figs.~1(c) and (d), the qubit is placed at the end of the control line, while the JQF is located a distance $d$ away from the qubit~\cite{Koshino2012,Hoi2015}.
The qubit and JQF are capacitively coupled to the control line.
Here, we consider that the JQF frequency $\omega_{\rm f}$ is set to be identical to the qubit frequency $\omega_{\rm q}$.
The resonant interaction mediated by the control line induces two cooperative effects depending on the distance: an energy-exchange interaction and a correlated decay~\cite{Chang2012,Lalumiere2013}.
Denoting the external coupling rates of the qubit and JQF by $\gamma_{\rm ex}^{\rm q}$ and $\gamma_{\rm ex}^{\rm f}$, the master equation in the rotating frame at the qubit frequency is given by
$\dot{\hat{\rho}}=-i/\hbar[\hat{H}_{\rm eff},\hat{\rho}]+\sum_{n,m={\rm q,f}}\gamma_{\rm ex}^{nm}[\hat{\sigma}_m\hat{\rho}\hat{\sigma}_n^\dag-(\hat{\sigma}_n^\dag\hat{\sigma}_m\hat{\rho}+\hat{\rho}\hat{\sigma}_n^\dag\hat{\sigma}_m)/2]$, where $\hat{\sigma}_{\rm q}$ and $\hat{\sigma}_{\rm f}$ are the lowering operators of the qubit and the JQF, respectively.
The energy-exchange interaction is described as $\hat{H}_{\rm eff}=\hbar J(\hat{\sigma}_{\rm q}^\dag\hat{\sigma}_{\rm f}+\hat{\sigma}_{\rm q}\hat{\sigma}_{\rm f}^\dag)$, where $J=\sqrt{\gamma_{\rm ex}^{\rm q}\gamma_{\rm ex}^{\rm f}}\sin\left(2\pi d/\lambda_{\rm q}\right)/2$ is the coupling strength and $\lambda_{\rm q}$ is the wavelength at the qubit frequency.
The decay terms are described with the individual decay rate $\gamma_{\rm ex}^{nn}=\gamma_{\rm ex}^n$ and the correlated decay rate $\gamma_{\rm ex}^{\rm qf}=\gamma_{\rm ex}^{\rm fq}=\sqrt{\gamma_{\rm ex}^{\rm q}\gamma_{\rm ex}^{\rm f}}\cos\left(2\pi d/\lambda_{\rm q}\right)$.
To avoid the qubit hybridizing with the JQF~($J=0$) and to maximize the correlated decay~($\gamma_{\rm ex}^{\rm qf}=\gamma_{\rm ex}^{\rm fq}=-\sqrt{\gamma_{\rm ex}^{\rm q}\gamma_{\rm ex}^{\rm f}}$), the distance $d$ is set to half the qubit wavelength~($d=\lambda_{\rm q}/2$).
Then, the correlated-decay modes are diagonalized by a dark mode  [$\hat{\sigma}_{\rm D}={\mathcal N}(\sqrt{\gamma_{\rm ex}^{\rm f}}\:\hat{\sigma}_{\rm q}+\sqrt{\gamma_{\rm ex}^{\rm q}}\:\hat{\sigma}_{\rm f})$] with decay rate $\gamma_{\rm ex}^{\rm D}=0$ and a bright mode [$\hat{\sigma}_{\rm B}={\mathcal N}(-\sqrt{\gamma_{\rm ex}^{\rm q}}\:\hat{\sigma}_{\rm q}+\sqrt{\gamma_{\rm ex}^{\rm f}}\:\hat{\sigma}_{\rm f})$] with decay rate $\gamma_{\rm ex}^{\rm B}=\gamma_{\rm ex}^{\rm q}+\gamma_{\rm ex}^{\rm f}$, where $\mathcal N$ is the normalization factor~\cite{Lalumiere2013}.
By engineering the system such that $\gamma_{\rm ex}^{\rm f}\gg\gamma_{\rm ex}^{\rm q}$, the excited state of the qubit is close to the dark state~($\hat{\sigma}_{\rm q}\approx\hat{\sigma_{\rm D}}$), suppressing its radiative decay.

In the experiment, a superconducting transmon qubit coupled to a coplanar-waveguide control line is fabricated on a silicon substrate~\cite{Koch2007}, as shown in Figs.~1(c).
The resonance frequency, anharmonicity, and external coupling rate of the qubit are $\omega_{\rm q}/2\pi=8.002$~GHz, $ \alpha_{\rm q} /2\pi= -398 $~MHz, and $ \gamma_{\rm ex}^{\rm q}/2\pi = 123 $~kHz, respectively.
The state of the qubit is dispersively read out via a resonator with resonance frequency $ \omega _ {\rm r} /2\pi=10.156$~GHz, external coupling rate $\kappa_{\rm ex}/2\pi = 2.16 $~MHz, and state-dependent dispersive frequency shift $2\chi/2 \pi = -1.87 $~MHz.
A  Purcell filter is used to prevent the Purcell decay from limiting the qubit relaxation time~\cite{Jeffrey2014,Sete2015}.
An ancillary transmon qubit acting as the JQF is strongly coupled to the control line with external coupling rate $ \gamma_{\rm ex}^{\rm f}/2\pi = 112$~MHz.
The distance between the qubit and JQF is designed to be half the qubit wavelength~($d\approx7.5$~mm).
The JQF resonance frequency $\omega_{\rm f}/2\pi$ is tunable between 6.3~GHz and 8.5~GHz with a static magnetic field, which enables us to investigate the behavior of the qubit with and without the JQF in a single device: when the JQF is far detuned from the qubit, the qubit behaves as if the JQF does not exist.
The anharmonicity and intrinsic loss rate of the JQF are $\alpha_{\rm f}/2\pi=-387$~MHz and $\gamma_{\rm in}^{\rm f}/2\pi=3$~MHz, respectively.

First, the JQF is characterized by measuring its reflection spectrum via the control line.
In Figs.~1(e) and (f), the amplitude and phase of the reflection coefficient as a function of the probe frequency are shown for different probe powers.
At a smaller probe power of $-146$~dBm, the JQF spectrum is in the over-coupling regime, where the external coupling rate is much larger than the intrinsic loss rate, i.e.\ $\gamma_{\rm ex}^{\rm f}\gg\gamma_{\rm in}^{\rm f}$.
The over-coupling regime of the JQF is required for a perfect reflection of single photons emitted from the qubit~\cite{Astafiev2010,Hoi2011}. 
The JQF transition starts to saturate around the single-photon power level~($\approx-120$~dBm).
The second dip around 7.8~GHz corresponds to the two-photon transition between the ground and second excited states.
At a stronger probe power of $-100$~dBm, the JQF does not affect the reflection coefficient due to it being saturated, which is an essential property for allowing the qubit control field to be transmitted through the JQF.

\begin{figure}[!t]
\begin{center}
  \includegraphics[width=80mm]{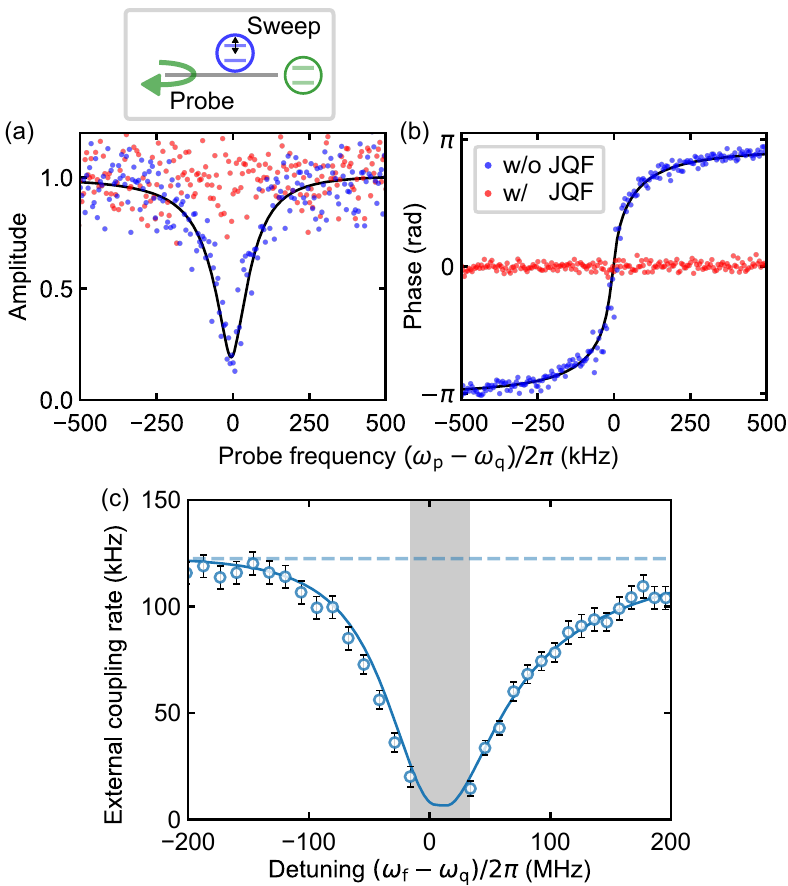} 
\caption{
Decoupling of the qubit from the control line by the JQF.
(a)~Amplitude and (b)~phase of the reflection spectra of the qubit measured via the control line.
The spectra are measured with a probe power of $-166$~dBm, which gives a linear response of the qubit.
The red and blue dots are the experimental results with and without the JQF, respectively. 
The black line is the theoretical fit to the data without the JQF.
(c)~External coupling rate of the qubit as a function of the JQF-qubit detuning.
The circles and the solid line are the experimental and numerically simulated results, respectively. 
The horizontal dashed line is the data in the absence of the JQF.
The gray-shaded area depicts the under-coupling regime, where it is hard to obtain a good fit because of the low signal-to-noise ratio. 
} 
  \label{fig2} 
\end{center}
\end{figure}

\begin{figure}[!t]
\begin{center}
  \includegraphics[width=80mm]{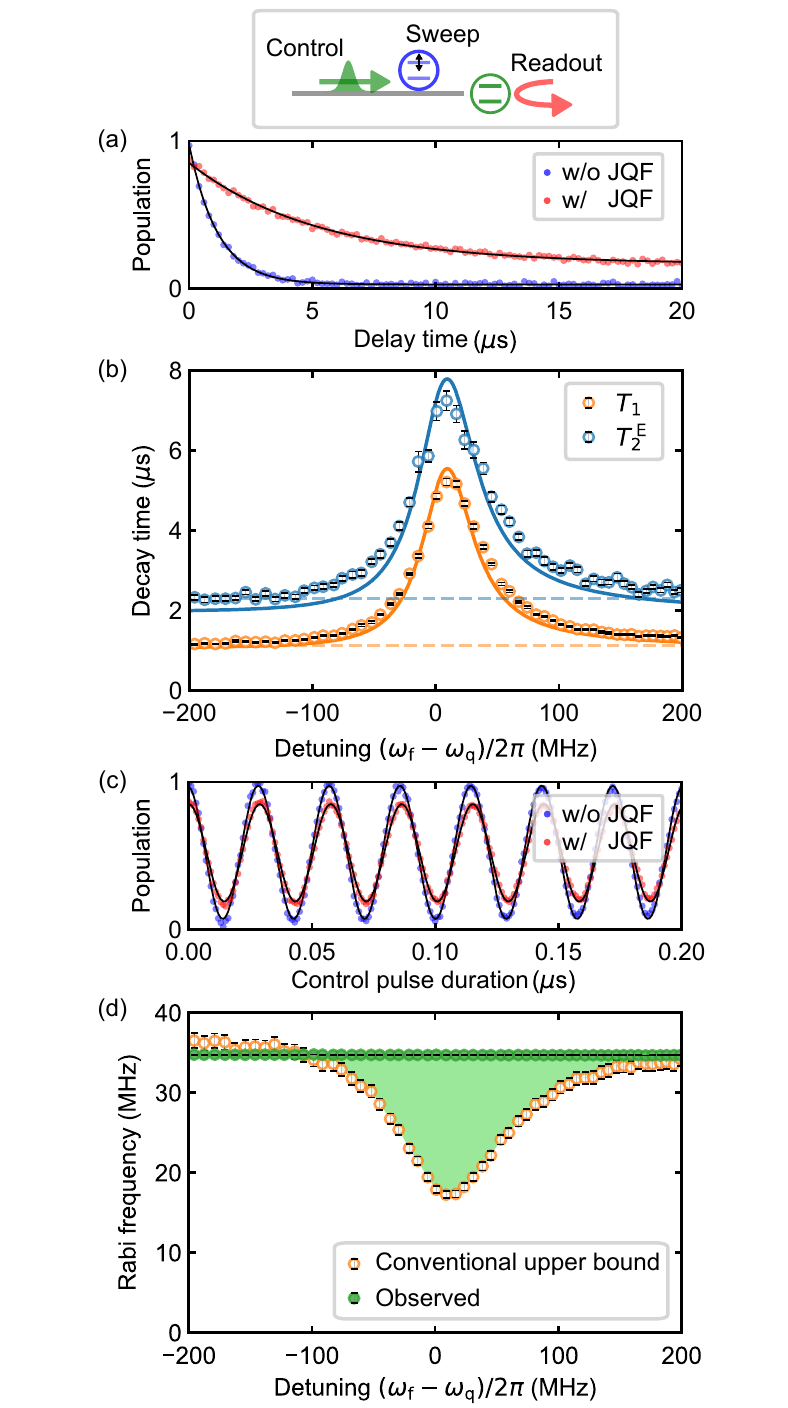} 
\caption{
Breaking the trade-off between the fast control and lifetime of the qubit with the JQF.
The red and blue dots are the experimental results with and without the JQF, respectively.
The black lines are the theoretical fits.
(a)~Qubit population in the excited state as a function of the delay time between the $\pi$ pulse and the readout.
(b)~Decay times as a function of the JQF-qubit detuning.
The orange and cyan circles depict the relaxation time~($T_1$) and the Hahn-echo coherence time~($T_2^{\rm E}$), respectively.
The solid lines are the numerically simulated results.
The horizontal dashed lines are those in the absence of the JQF.
(c)~Qubit population in the excited state as a function of the duration of a square control pulse with Gaussian edges.
(d)~Rabi frequency as a function of the JQF-qubit detuning.
The green circles are the observed Rabi frequency in the presence of the JQF, and the orange circles are the conventional upper bound of the Rabi frequency for the given drive power, which is calculated as $2\sqrt{\dot{n}/T_1}$. 
} 
  \label{fig3} 
\end{center}
\end{figure}

Next, the qubit reflection spectra are measured via the control line with different detunings between the JQF and the qubit~[see Figs.~2(a) and (b)].
The qubit spectrum without the JQF is found to be in the over-coupling regime to the control line.
In contrast, when the JQF is on resonance with the qubit, the signature of the qubit transition disappears, indicating that the qubit is decoupled from the control line. 
In Fig.~2(c), the qubit external coupling rate is shown as a function of the JQF-qubit detuning.
The full frequency bandwidth at half maximum of the suppression spectrum is found to be about $130$~MHz, which roughly coincides with the external coupling rate of the JQF.
The asymmetry of the suppression spectrum is explained by the non-ideal distance between the qubit and the JQF, i.e.\ $d=0.526\lambda_{\rm q}$, in the actual device.
Interestingly, for the non-ideal distance, we find that the external coupling rate of the qubit can be maximally suppressed at a detuning of $(\omega_{\rm f}-\omega_{\rm q})/2\pi=9$~MHz.

To characterize the dynamics of the qubit in the presence of the JQF, time-domain measurements are performed for different JQF-qubit detunings.
The qubit population in the excited state is obtained from the averaged quadrature amplitude of the readout pulse.
Note that the averaged amplitude is normalized by taking into account the thermal population of the qubit for each detuning~[see sections S6 in \cite{supple}].

As shown in Fig.~3(a), we measure the qubit relaxation time by observing the qubit population as a function of the delay time between the $\pi$ pulse and the qubit readout.
When the JQF is nearly resonant with the qubit, the qubit shows an exponential decay with a longer relaxation time than that without the JQF.
Furthermore,the thermal population of the qubit is increased from 2.8\% to 16.2\%, which indicates that the effective temperature of the intrinsic loss channel of the qubit is higher than that of the control line~\cite{Jin2015}.
The relaxation time~($T_1$) and Hahn-echo coherence time~($T_2^\mathrm{E}$) of the qubit as a function of the JQF-qubit detuning are shown in Fig.~3(b).
Both the relaxation and coherence times are enhanced when the JQF is nearly on resonance with the qubit.
The enhancement ratios are mainly limited by the intrinsic energy relaxation and thermal population of the qubit.

Furthermore, Rabi oscillations of the qubit are observed when applying a stationary control field with a photon flux of $\dot{n}=1.5\times10^{10}\:{\rm s}^{-1}$, which corresponds to $-101$~dBm.
As shown in Fig.~3(c), the Rabi oscillations are not affected by the presence of the JQF except for the oscillation visibility which is decreased due to the thermal excitation caused by the intrinsic loss channel of the qubit.
The observed Rabi frequency as a function of the JQF-qubit detuning is shown with the green circles in Fig.~3(d).
Due to the saturation of the JQF by the strong control field, the Rabi frequency is found to be constant and does not depend on the detuning.

To further study the trade-off in qubit control, we define the conventional upper bound of the Rabi frequency of the qubit without employing the JQF as $2\sqrt{\dot{n}/T_1}$.
This is because a Rabi frequency $\Omega_{\rm q}$ with a fixed external coupling never exceeds the upper bound, as $\Omega_{\rm q}=2\sqrt{\dot{n}\gamma_{\rm ex}^{\rm q}} \leq 2\sqrt{\dot{n}/T_1}$.
The upper bound can be achieved in a conventional control-line setup only when the internal loss, pure dephasing and thermal excitation of the qubit are negligible.
As shown in Fig.~3(d), the observed Rabi frequency with the JQF exceeds this upper bound~[indicated by the shadowed area in Fig.~3(d)], which demonstrates that we break the trade-off in qubit control.

\begin{figure*}[!t]
\begin{center}
  \includegraphics[width=160mm]{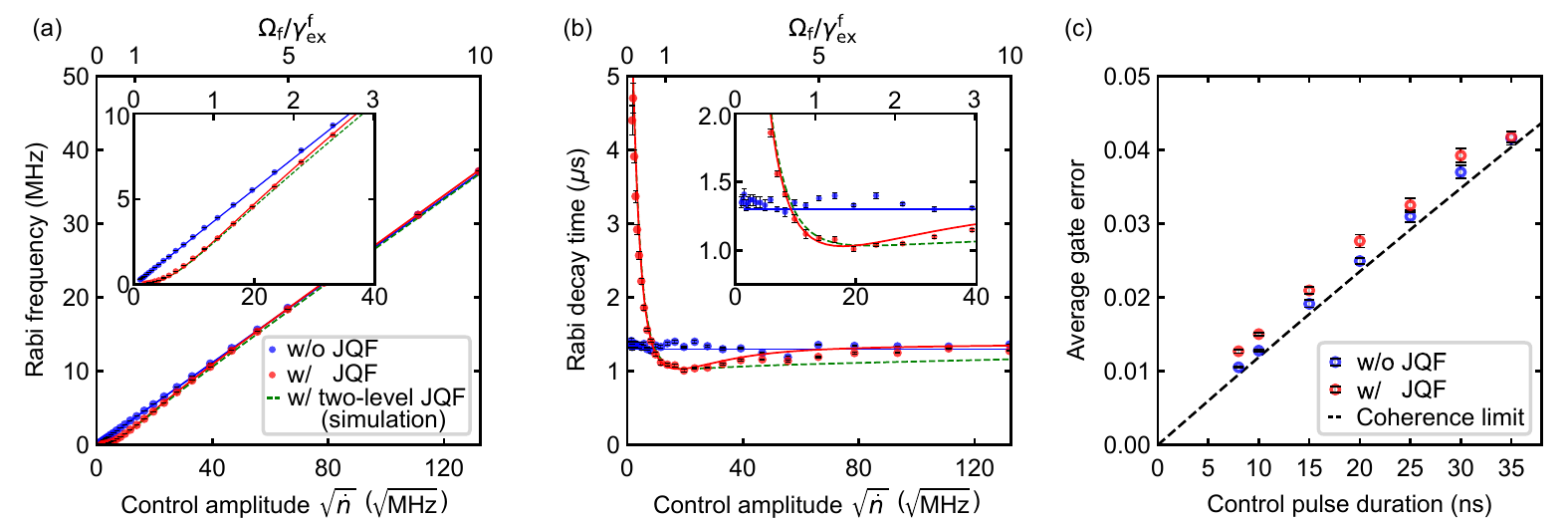} 
\caption{
Rabi control of the qubit with the JQF.
(a)~Rabi frequency and (b)~Rabi decay time of the qubit as a function of the control amplitude represented as the square root of the photon flux $\sqrt{\dot{n}}$.
The top axes show the corresponding Rabi frequency of the JQF $\Omega_{\rm f}=2\sqrt{\gamma_{\rm ex}^{\rm f}\dot{n}}$ normalized by its external decay rate $\gamma_{\rm ex}^{\rm f}$.
The red and blue data are the experimental results with and without the JQF, respectively.
The red and blue solid lines depict numerical results with and without a transmon JQF, while the green dashed lines show simulation results using a two-level JQF. 
(c)~Average gate error of the qubit as a function of the control pulse duration, obtained by randomized benchmarking.
A Gaussian pulse is used for the qubit control, and the pulse interval is set to be twice the pulse duration.
The black dashed line depicts the coherence limit.
} 
  \label{fig4} 
\end{center}
\end{figure*}

To investigate if the JQF has negative effects on our ability to control the qubit, the Rabi frequency and Rabi decay time of the qubit are measured as a function of the control amplitude~[Figs.~4(a) and (b)].
As expected, the Rabi frequency in the absence of the JQF increases linearly with the control amplitude, and the Rabi decay time is constant.
In the presence of the JQF, on the other hand, in the region where the JQF is not completely saturated~($\Omega_{\rm f}/\gamma_{\rm ex}^{\rm f}\approx1$), the Rabi frequency is smaller and the decay time is shorter than that without the JQF.
However, in the limit of large control amplitudes~($\Omega_{\rm f}/\gamma_{\rm ex}^{\rm f}\gg1$), the results with and without the JQF become identical.

In Figs.~4(a) and (b), we compare the experimental results with the numerical ones which we calculate by replacing the transmon JQF with a two-level JQF with the same parameters.
Unlike with the transmon JQF, the Rabi decay time of the qubit with the two-level JQF is calculated to be shorter than that in the absence of the JQF, even when the JQF is nearly completely saturated by the control field~($\Omega_{\rm f}/\gamma_{\rm ex}^{\rm f}\gg1$).
This is because the saturated two-level JQF is still resonantly coupled to the qubit, providing an additional decay channel of the qubit.
On the other hand, since the transmon JQF, once excited to its higher levels, becomes decoupled from the qubit due to its anharmonicity, it no longer affects the Rabi oscillations of the qubit.
From numerical simulations, we find that the minimum error per qubit Rabi cycle is achieved when the anharmonicity of a transmon JQF almost equals its external coupling rate ($|\alpha_{\rm f}|\approx\gamma_{\rm ex}^{\rm f}$), which agrees with our experimental findings~(see section S4 in \cite{supple}).

When the qubit is sequentially controlled, the JQF is expected to be saturated during each gate and not to significantly affect gate fidelity.
The average gate error of the Clifford gates on the qubit is measured by using randomized benchmarking~\cite{Magesan2012}.
In Fig.~4(c), the average gate errors with and without the JQF are shown as a function of the control pulse duration.
We confirm that the average gate errors in both cases are close to the coherence limit and that the JQF is not harmful to the qubit control.
The small increase in the gate error with the JQF can be explained by the additional decay of the qubit due to the incomplete saturation of the JQF at the beginning and end of the control pulse. 
Note that the coherence limit is mainly determined by the external coupling rate of the qubit.

In conclusion, we successfully resolved the trade-off in qubit control by implementing a Josephson quantum filter~(JQF) to the control line of a qubit.
We experimentally confirmed that the JQF suppresses the qubit radiative decay, while it does not significantly reduce the Rabi frequency and the gate fidelity of the qubit.
The device could be useful in the realization of a large-scale superconducting qubit system by reducing the heating of the qubit environment and the crosstalk between qubits.
More generally, our experiments show that a nonlinear element acts as a power-dependent variable boundary condition for microwave modes, which can be applied to other types of parametric control, such as two-qubit gates, single-photon generation, or active cooling of quantum systems.

\textbf{Acknowledgments:}
We acknowledge fruitful discussions with A. Eddins, J. M. Kreikebaum, and K. O'Brien.
This work was supported in part by UTokyo ALPS, JSPS Fellows (No.~18J13084), JSPS KAKENHI (No.~19K03684 and No.~26220601), JST ERATO (No.~JPMJER1601), and MEXT Q-LEAP (No.~JPMXS0118068682).
D.L.-Q. and A.F.v.L. are International Research Fellows of JSPS.

\textbf{Author contributions:}
S.K. conceived the concept, designed and fabricated the sample.
K.K and S.K. provided theoretical models.
S.K. and D.L.-Q. performed the experiments.
D.L.-Q. and S.K. set up the electronics for measurement.
S.K. analyzed the results and wrote the manuscript with feedback from all the coauthors. 
Y.N. supervised the project.

\bibliography{mybib}

\clearpage
\begin{widetext}
\setcounter{equation}{0}
\setcounter{figure}{0}
\setcounter{table}{0}
\renewcommand{\theequation}{S\arabic{equation}}
\renewcommand{\thefigure}{S\arabic{figure}}
\renewcommand{\thetable}{S\arabic{table}}

\section*{\textbf{\normalsize Supplementary Materials}}


\section*{S1. Sample information}
An optical microscope image of a nominally identical sample to the one used in this work is shown in Figs.~S1(a)--(f).
Three qubits are dispersively coupled to their respective readout resonators, which are connected to a readout line via a Purcell filter.
The transmon qubit, which is coupled to a control coplanar waveguide, is used as a data qubit in the main experiments~[see Figs.~S1(d) and (e)].
A Josephson quantum filter~(JQF), which is a frequency-tunable transmon qubit with a capacitor composed of a single electrode and the ground plane, is strongly coupled to the control line approximately half a wavelength apart from the data qubit~[see Figs.~S1(b) and (c)].

\begin{figure}[!b]
\begin{center}
  \includegraphics[width=160mm]{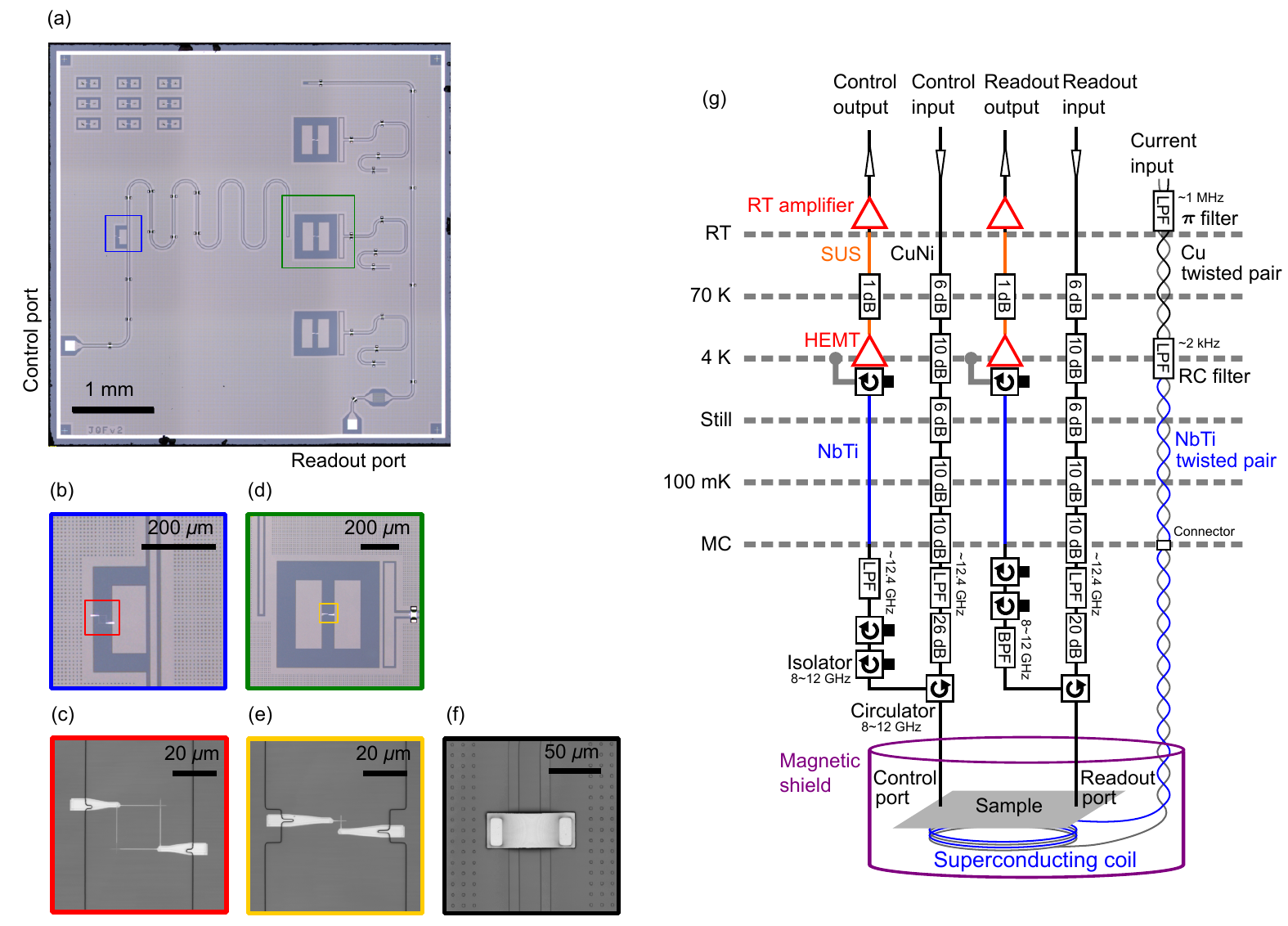} 
\caption{Sample and experimental setup.
(a)~Optical microscope image of the entire sample.
(b)~JQF and and (c)~its asymmetric SQUID.
(d)~Data qubit and (e)~its single Josephson junction.
(f)~An air-bridge.
(g)~Schematic of the experimental setup.
}
\end{center}
\end{figure}

To fabricate the superconducting circuit, a 200-nm thick Nb film on a high-resistivity ($>$10~${\rm k\Omega}$-cm) silicon substrate is patterned with photolithography followed by ${\rm CF_4}$ plasma etching to form the larger structures, such as the transmon pads, JQF pad, readout resonators, Purcell filter, and coplanar waveguides.
The magnetic-flux-trapping holes are also fabricated on the ground plane in this step.
Following $\rm O_2$ plasma ashing and HF cleaning, the Al/Al${\rm O}_x$/Al Josephson junctions are fabricated with electron-beam lithography followed by electron-beam evaporation using a bridgeless process.
While the fixed frequency data qubit contains a single Josephson junction, the JQF contains an asymmetric SQUID in order to be able to tune its frequency to be resonant with the data qubit~\cite{Hutchings2017}.
The 300-nm thick Al air-bridges with heights of about 3~$\mu$m~[see Fig.~S1(f)] are fabricated with a four-step process: photolithography followed by electron-beam evaporation and another photolithography followed by Al etching. 
Together with the Al air-bridges, a frame and bonding pads made of Al are fabricated to make good contacts with Al bonding wires. 
After dicing, the device is carefully cleaned with NMP and with $\rm O_2$ plasma ashing.

As listed in Table~S1, we determine the system parameters by comparing the experimental data with theoretical models in Sec.~S3.

\begin{table}[!b]
\begin{center}
\caption{
System parameters. Note that $\lambda_{\rm q}$~($\approx15$~mm) is the wavelength at the qubit frequency.
}
  \begin{tabular}{lc} \\ 
\qquad \qquad Qubit & Value \\ \hline \hline
Resonance frequnecy, $\omega_{\rm q}/2\pi$  & 8.002 GHz \\ \hline 
Anharmonicity, $\alpha_{\rm q}/2\pi$ & $-$0.398 GHz \\ \hline 
External coupling rate, $\gamma_{\rm ex}^{\rm q}/2\pi$  & 123 kHz \\ \hline 
Intrinsic loss rate, $\gamma_{\rm in}^{\rm q}/2\pi$ & 16 kHz \\ \hline
Pure dephasing rate, $\gamma_{\phi}^{\rm q}/2\pi$ & 6 kHz \\ \hline
Intrinsic thermal quanta, $n_{\rm th}^{\rm q}$ & 0.29 \\ \hline 
 & w/o JQF  \quad w/ JQF \\ \hline 
Relaxation time, $T_1$ & \quad1.1 $\mu$s \qquad  5.2 $\mu$s \\ \hline 
Coherence time, $T_2^{\rm E}$ & \quad 2.3 $\mu$s \qquad  7.3 $\mu$s \\ \hline 
Coherence time, $T_2^*$  & \quad 1.7 $\mu$s \qquad  3.8 $\mu$s \\ \hline 
Thermal population, $p_{\rm th}^{\rm q}$ & \: 0.028 \qquad 0.16\\ \hline 
\\ 
\qquad \qquad JQF  & Value \\ \hline \hline
Resonance frequnecy, $\omega_{\rm f}/2\pi$  & 6.3$-$8.5 GHz \\ \hline 
Anharmonicity, $\alpha_{\rm f}/2\pi$ & $-$0.387 GHz \\ \hline 
External coupling rate, $\gamma_{\rm ex}^{\rm f}/2\pi$  & 113 MHz \\ \hline 
Intrinsic loss rate, $\gamma_{\rm in}^{\rm f}/2\pi$ & 3 MHz \\ \hline
Pure dephasing rate, $\gamma_{\phi}^{\rm f}/2\pi$ & 0 \\ \hline
Intrinsic thermal quanta, $n_{\rm th}^{\rm f}$ & 0 \\ \hline 
Separation between qubit and JQF, $d$ & $0.526\lambda_{\rm q}$ \\ \hline 
\\ 
\qquad \qquad Readout resonator  & Value \\ \hline \hline
Resonance frequency, $\omega_{\rm r}/2\pi$  & 10.1564 GHz \\ \hline 
External coupling rate, $\kappa_{\rm ex}/2\pi$  & 2.152 MHz \\ \hline 
Intrinsic loss rate, $\kappa_{\rm in}/2\pi$ & 0.015 MHz \\ \hline
Dispersive frequency shift, $2\chi/2\pi$ & 1.870 MHz \\ \hline
  \end{tabular}
\end{center}
\end{table}

\section*{S2. Experimental setup}
The experimental setup is shown in Fig.~S1(g).
The superconducting circuit is mounted in a sample holder placed in a magnetic shield at the base temperature stage ($\sim 50$~mK) of a dilution refrigerator.
A magnetic field is applied to the SQUID of the JQF by a superconducting coil placed around the chip.
Radio-frequency noise through the flux-bias line is removed by low-pass filters: a $\pi$ filter at room temperature and an RC filter at the 4~K stage.
The microwave input line is highly-attenuated and filtered to decrease noise from room temperature.
A cryogenic HEMT amplifier and a room-temperature low-noise amplifier are installed in the microwave output line to measure weak microwave signals.
Several circulators and filters are installed in the output line to reduce the backward noise from the amplifiers.

The readout resonator is characterized via the readout line using reflection measurements in the frequency domain.
The qubit is characterized using two-tone spectroscopy.
Furthermore, the qubit and JQF are further characterized via the control line by the reflection measurements in the frequency domain.
Then, the qubit state is controlled by applying resonant microwave pulses via the control line, and is read out by measuring the dispersive shift of the readout resonator via the readout line in the time domain.
We study the effect of the JQF on the qubit control by measuring the relaxation and Rabi oscillation of the qubit in both cases when the JQF is in or out of resonance with the qubit.

The microwave pulses for control and readout are generated by mixing low-frequency pulse signals with continuous microwaves. 
The low-frequency signals are generated at 1-GHz sampling rates with digital-analog converters. 
The readout pulses are down-converted at the mixer using a continuous microwave signal phase-locked with the one used for the pulse generation and are measured at a 1-GHz sampling rate by an analog-digital converter.

\section*{S3. Theoretical model}
To numerically simulate the experimental data, we generalize the theoretical model discussed in the reference~\cite{Koshino2020}, i.e.\ we consider a system where two transmon qubits (a data qubit and a JQF) with different resonance frequencies are coupled to a semi-infinite waveguide, as shown in Fig.~S2(a).
In this system, a propagating field interacts with each qubit twice at different points before and after the reflection by the open end.
Therefore, it can be modeled as a system where two qubits are coupled at two different points to an infinite waveguide containing only right-propagating modes, as shown in Fig.~S2(b).
The qubit (JQF) is coupled to the waveguide at positions $-r_{\rm q}$ and $r_{\rm q}$ ($-r_{\rm f}$ and $r_{\rm f}$) with external coupling rate of $\gamma_{\rm ex}^{\rm q}/4$ ($\gamma_{\rm ex}^{\rm f}/4$), where it is assumed that $r_{\rm q} \leq r_{\rm f}$.

\subsection*{Cooperative effects mediated by a waveguide}
We derive the cooperative effects on the qubit and the JQF mediated by the waveguide.
Setting $\hbar=v=1$, where $v$ is the velocity of the microwaves, the total Hamiltonian under the rotating wave approximation is given by
\begin{equation}
\label{Htot}
\hat{H} = \hat{H}_{\rm sys}+\int_{-\infty}^\infty dk\:k \hat{a}_k^\dag \hat{a}_k+\sum_{n={\rm q,f}}\int_{-\infty}^{\infty} \frac{dk}{\sqrt{2\pi}}\sqrt{\frac{\gamma_{\rm ex}^n}{4}}\left(\hat{b}_n^\dag  \hat{a}_k e^{-ikr_n} + \hat{b}_n\hat{a}_k^\dag e^{+ikr_n}+\hat{b}_n^\dag  \hat{a}_k e^{+ikr_n} + \hat{b}_n\hat{a}_k^\dag e^{-ikr_n}\right),
\end{equation}
where $\hat{H}_{\rm sys} = \sum_{n={\rm q,f}}\:\omega_n \hat{b}_n^\dag\hat{b}_n + \frac{\alpha_n}{2}\hat{b}_n^{\dag2}\hat{b}_n^2$ is the system Hamiltonian of the two transmon qubits with resonance frequencies $\omega_n$ and anharmonicities $\alpha_n$, $\hat{b}_n$ is the annihilation operator of the transmon qubit $n$, and $\hat{a}_k$ is the annihilation operator of the right-propagating mode with wavenumber $k$.
The coupling strength of the propagating mode and the system is assumed not to depend on the wavenumber. 
Furthermore, it is assumed that the lower limit of $k$ integration is extended to $-\infty$ since the propagating mode with a negative frequency does not affect the dynamics due to the large detuning from the system frequencies.
The waveguide mode operator in the wavenumber representation, $\hat{a}_k$, is associated with that in the real-space representation, $\hat{a}_r$, as
\begin{equation}
\label{fif}
\hat{a}_k=  \int_{-\infty}^{\infty} \frac{dr}{\sqrt{2\pi}}\:\hat{a}_re^{-ikr}
\quad{\rm and}\quad 
\hat{a}_r =  \int_{-\infty}^{\infty} \frac{dk}{\sqrt{2\pi}}\:\hat{a}_ke^{ikr}.
\end{equation}
They obey the following commutation relations: $\left[\hat{a}_k,\hat{a}_{k'}^\dag\right] = \delta(k-k')$ and $\left[\hat{a}_r,\hat{a}_{r'}^\dag\right] = \delta(r-r')$.

\begin{figure}[!t]
\begin{center}
  \includegraphics[width=110mm]{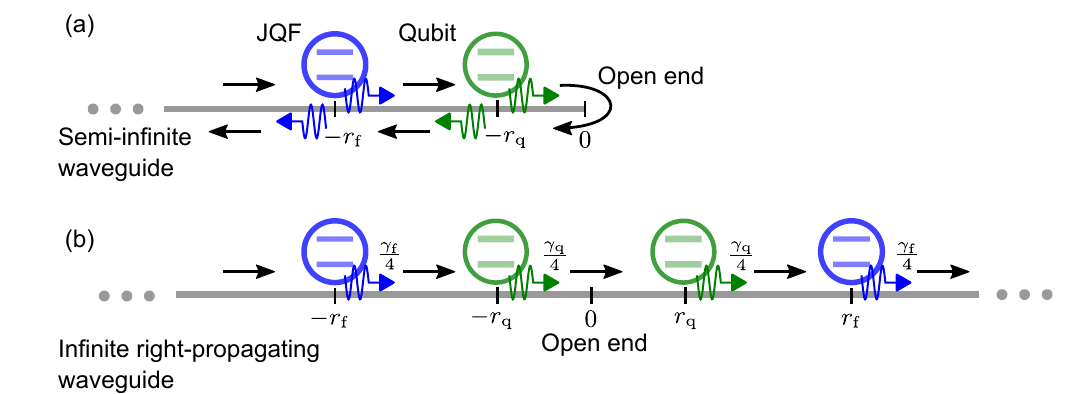} 
\caption{Schematic of the theoretical model.
(a)~System where a qubit and a JQF are coupled to a semi-infinite waveguide.
(b)~System where a qubit and a JQF are coupled twice to the right-propagating modes of an infinite waveguide.
}
\end{center}
\end{figure}

From Eq.~(\ref{Htot}), the Heisenberg equation for $\hat{a}_k$ is given by
\begin{equation}
\frac{d\hat{a}_k}{dt} =  -ik\hat{a}_k-\frac{i}{\sqrt{2\pi}}\sum_{n={\rm q,f}}\sqrt{\frac{\gamma_{\rm ex}^n}{4}}\left(\hat{b}_n e^{+ikr_n}+\hat{b}_n e^{-ikr_n}\right),
\end{equation}
which can be formally solved as
\begin{equation}
\label{rak}
\hat{a}_k= \hat{a}_k(0) e^{-ikt} -\frac{i}{\sqrt{2\pi}}\sum_{n={\rm q,f}}\sqrt{\frac{\gamma_{\rm ex}^n}{4}}\int_0^{t}dt'\left(\hat{b}_n(t') e^{-ik(-r_n-t'+t)}+\hat{b}_n(t') e^{-ik(r_n-t'+t)}\right),
\end{equation}
where $\hat{a}_k(0)$ is the right-propagating annihilation operator with wavenumber $k$ at the initial time $t=0$ and $\hat{b}_n(t')$ is the annihilation operator of the transmon qubit at time $t'$.
Note that operators at time $t$ are written in a simple form hereinafter, e.g.\ $\hat{a}_k=\hat{a}_k(t)$.
Multiplying Eq.~(\ref{rak}) by $e^{ikr}/\sqrt{2\pi}$ and integrating from $-\infty$ to $\infty$ with respect to $k$, we obtain the input-output relation of the right-propagating mode in the real-space representation as
\begin{equation}
\label{ior}
\hat{a}_r =  \hat{a}_{r-t}(0) -i\sum_{n={\rm q,f}}\sqrt{\frac{\gamma_{\rm ex}^n}{4}}\left\{\Theta_{r\in(-r_n,-r_n+t)}\:\hat{b}_n\!\left(t-r-r_n\right) + \Theta_{r\in(r_n,r_n+t)}\:\hat{b}_n\!\left(t-r+r_n\right)\right\},
\end{equation}
where $\Theta$ is a product of Heaviside step functions, $\Theta_{r\in(a,b)} = \theta(r-a)\theta(b-r)$.

Using Eqs. (\ref{Htot}) and (\ref{fif}), the time evolution of an arbitrary operator $\hat{O}$ supported by the two-qubit subspace is described in the Heisenberg picture as 
\begin{equation}
\label{dOdt}
\frac{d\hat{O}}{dt} 
=  i \left[\hat{H}_{\rm sys},\hat{O}\right]+i\sum_{n={\rm q,f}} \sqrt{\frac{\gamma_{\rm ex}^n}{4}}\left(\left[\hat{b}_n^\dag ,\hat{O}\right](\hat{a}_{-r_n}+\hat{a}_{r_n})+(\hat{a}_{-r_n}^\dag+\hat{a}_{r_n}^\dag)\left[\hat{b}_n,\hat{O}\right]\right).
\end{equation}
By substituting Eq. (\ref{ior}) into Eq.~(\ref{dOdt}), it can be rewritten as  
\begin{equation}
\label{dOdt2}
\begin{aligned}
\frac{d\hat{O}}{dt} 
= \:&i \left[\hat{H}_{\rm sys},\hat{O}\right]+i\sum_{n={\rm q,f}} \sqrt{\gamma_{\rm ex}^n}\left(\left[\hat{b}_n^\dag ,\hat{O}\right]\frac{\hat{a}_{-r_n-t}(0)+\hat{a}_{r_n-t}(0)}{2}+\frac{\hat{a}_{-r_n-t}^\dag(0)+\hat{a}_{r_n-t}^\dag(0)}{2}\left[\hat{b}_n,\hat{O}\right]\right)\\
&+\sum_{n={\rm q,f}}\frac{\gamma_{\rm ex}^n}{2}\left(\left[\hat{b}_n^\dag ,\hat{O}\right]\frac{\hat{b}_n + \hat{b}_n(t-2r_n)}{2}-\frac{\hat{b}_n^\dag + \hat{b}_n^\dag(t-2r_n)}{2} \left[\hat{b}_n,\hat{O}\right]\right)\\
&+\sum_{n={\rm q,f}}\frac{\sqrt{\gamma_{\rm ex}^{\rm q}\gamma_{\rm ex}^{\rm f}}}{2}\left(\left[\hat{b}_n^\dag ,\hat{O}\right]\frac{\hat{b}_{\bar{n}}(t+r_{\rm q}-r_{\rm f})+\hat{b}_{\bar{n}}(t-r_{\rm q}-r_{\rm f})}{2}-\frac{\hat{b}_{\bar{n}}^\dag(t+r_{\rm q}-r_{\rm f})+\hat{b}_{\bar{n}}^\dag(t-r_{\rm q}-r_{\rm f})}{2}\left[\hat{b}_n,\hat{O}\right]\right),
\end{aligned}
\end{equation}
where $\bar{n} = {\rm f}\:(\bar{n} ={\rm q})$ if $n={\rm q} \:(n={\rm f})$.

Next, we employ the free-evolution approximation, $\hat{b}_n(t-x)\approx\hat{b}_n e^{i\omega_nx}$, which is valid when the time delay is much shorter than the typical time scales of dynamics of the qubits, such as the radiative lifetime.
Furthermore, the initial state of the propagating mode is assumed to be separable from the system of the qubit and JQF and to be either in the vacuum state or a coherent state.
Since a coherent state, which generally includes the vacuum state, is an eigenstate of the annihilation operator with its eigenvalue being a coherent amplitude,
the annihilation operator of the propagating mode at the initial time $t=0$ can be replaced with its complex amplitude, e.g.\ $\hat{a}_{-t}(0) \rightarrow \sqrt{\dot{n}}\:e^{-i\omega t}$, where $\dot{n}$ is the photon flux of the coherent drive field and $\omega$ is the drive frequency.
Then, Eq.~(\ref{dOdt2}) is rewritten as 
\begin{equation}
\label{dOdt3}
\begin{aligned}
\frac{d\hat{O}}{dt} 
=  \: & i \left[\hat{H}_{\rm sys},\hat{O}\right]+i\sum_{n={\rm q,f}} \sqrt{\gamma_{\rm ex}^n\dot{n}}\:\cos(\omega r_n)\left[\hat{b}_n^\dag e^{-i\omega t}+\hat{b}_n e^{i\omega t},\hat{O}\right]\\
&+\sum_{n={\rm q,f}}\frac{\gamma_{\rm ex}^n\cos(\omega_nr_n)}{2}\left(\left[\hat{b}_n^\dag ,\hat{O}\right]\hat{b}_ne^{i\omega_nr_n}-\hat{b}_n^\dag \left[\hat{b}_n,\hat{O}\right]e^{-i\omega_nr_n}\right)\\
&+\sum_{n={\rm q,f}}\frac{\sqrt{\gamma_{\rm ex}^{\rm q}\gamma_{\rm ex}^{\rm f}}\:\cos(\omega_{\bar{n}}r_{\rm q})}{2}\left(\left[\hat{b}_n^\dag ,\hat{O}\right]\hat{b}_{\bar{n}}e^{i\omega_{\bar{n}}r_{\rm f}}-\hat{b}_{\bar{n}}^\dag e^{-i\omega_{\bar{n}}r_{\rm f}}\left[\hat{b}_n,\hat{O}\right]\right).
\end{aligned}
\end{equation}
The second term on the right-hand side is associated with the coherent-drive Hamiltonian of the qubit and the JQF, which is defined by
$
\hat{H}_{\rm drive} = \sum_{n={\rm q,f}} \sqrt{\gamma_{\rm ex}^n\dot{n}}\:\cos(\omega r_n)\left(\hat{b}_n^\dag e^{-i\omega t}+\hat{b}_n e^{i\omega t}\right)
$.

Using the cyclic invariance of trace operations, the time derivative of the expectation value $\langle\hat{O}\rangle$ is described in the Schr\"{o}dinger picture as
\begin{equation}
\label{dOdtHS}
\begin{aligned}
\frac{d\langle\hat{O}\rangle}{dt} = \:& {\rm Tr}\left[\frac{d\hat{O}}{dt}\:\hat{\rho}_{\rm T}(0)\right] ={\rm Tr}\left[\hat{O}(0)\:\frac{d\hat{\rho}_{\rm T}}{dt}\right]\\
= &-i\:{\rm Tr}\left[\hat{O}\:[\hat{H}_{\rm sys}+\hat{H}_{\rm drive},\:\hat{\rho}_{\rm T}]\right]\\
&+\sum_{n={\rm q,f}}\:{\rm Tr}\left[\hat{O}\frac{\gamma_{\rm ex}^n\cos(\omega_nr_n)}{2}\left\{\left(\hat{b}_n\hat{\rho}_{\rm T}\hat{b}_n^\dag-\hat{b}_n^\dag\hat{b}_n\hat{\rho}_{\rm T}\right)e^{i\omega_nr_n}+\left(\hat{b}_n\hat{\rho}_{\rm T}\hat{b}_n^\dag-\hat{\rho}_{\rm T}\hat{b}_n^\dag\hat{b}_n\right)e^{-i\omega_nr_n}\right\}\right]\\
&+\sum_{n={\rm q,f}}\:{\rm Tr}\left[\hat{O}\:\frac{\sqrt{\gamma_{\rm ex}^{\rm q}\gamma_{\rm ex}^{\rm f}}\:\cos(\omega_{\bar{n}}r_q)}{2}\left\{\left(\hat{b}_{\bar{n}}\hat{\rho}_{\rm T}\hat{b}_n^\dag- \hat{b}_n^\dag\hat{b}_{\bar{n}}\hat{\rho}_{\rm T} \right)e^{i\omega_{\bar{n}}r_{\rm f}}+\left(\hat{b}_n\hat{\rho}_{\rm T}\hat{b}_{\bar{n}}^\dag- \hat{\rho}_{\rm T}\hat{b}_{\bar{n}}^\dag\hat{b}_n \right)e^{-i\omega_{\bar{n}}r_{\rm f}}\right\}\right].
\end{aligned}
\end{equation}

The expectation value of the local operator $\hat{O}$ can be obtained from the reduced density matrix, in which the degrees of freedom of the propagating modes are traced out.
Therefore, by defining the density-matrix operator of the composite system of the qubit and JQF as $\rho = {\rm Tr}_{\rm p}[\rho_{\rm T}]$, where ${\rm Tr}_{\rm p}[\cdot]$ is the partial trace of the propagating mode,
the reduced master equation is derived from Eq.~(\ref{dOdtHS}) as
\begin{equation}
\begin{aligned}
\frac{d\hat{\rho}}{dt}= &-i\left[\hat{H}_{\rm sys}+\hat{H}_{\rm drive},\:\hat{\rho}\right]\\
&+\sum_{n={\rm q,f}}\frac{\gamma_{\rm ex}^n\cos(\omega_nr_n)}{2}\left[\left(\hat{b}_n\hat{\rho}\hat{b}_n^\dag-\hat{b}_n^\dag\hat{b}_n\hat{\rho}\right)e^{i\omega_nr_n}+\left(\hat{b}_n\hat{\rho}\hat{b}_n^\dag-\hat{\rho}\hat{b}_n^\dag\hat{b}_n\right)e^{-i\omega_nr_n}\right]\\
&+\sum_{n={\rm q,f}}\frac{\sqrt{\gamma_{\rm ex}^{\rm q}\gamma_{\rm ex}^{\rm f}}\:\cos(\omega_{\bar{n}}r_q)}{2}\left[\left(\hat{b}_{\bar{n}}\hat{\rho}\hat{b}_n^\dag- \hat{b}_n^\dag\hat{b}_{\bar{n}}\hat{\rho}\right)e^{i\omega_{\bar{n}}r_{\rm f}}+\left(\hat{b}_n\hat{\rho}\hat{b}_{\bar{n}}^\dag- \hat{\rho}\hat{b}_{\bar{n}}^\dag\hat{b}_n \right)e^{-i\omega_{\bar{n}}r_{\rm f}}\right].
\end{aligned}
\end{equation}
Then, the master equation can be rewritten in the Lindblad form as
\begin{equation}
\label{drhodt}
\begin{aligned}
\frac{d\hat{\rho}}{dt}= &-i\left[\hat{H}_{\rm sys}+\hat{H}_{\rm drive}+\hat{H}_{\rm Lamb}+\hat{H}_{\rm eff},\:\hat{\rho}\right]
+\sum_{n,m={\rm q,f}}\gamma_{\rm ex}^{nm}\:{\mathcal D}(\hat{b}_n,\hat{b}_m)\hat{\rho},
\end{aligned}
\end{equation}
where the Lamb-shift Hamiltonian is given by
$
\hat{H}_{\rm Lamb} = \sum_{n={\rm q,f}}\:\frac{\gamma_{\rm ex}^n}{4}\sin(2\omega_nr_n)\:\hat{b}_n^\dag\hat{b}_n
$.
The waveguide-mediated effective interaction Hamiltonian is given by
\begin{equation}
\hat{H}_{\rm eff} = J\:\hat{b}_{\rm q}^\dag\hat{b}_{\rm f} + J^* \:\hat{b}_{\rm f}^\dag \hat{b}_{\rm q},
\end{equation}
where
$
J= \sqrt{\gamma_{\rm ex}^{\rm q}\gamma_{\rm ex}^{\rm f}}/2\:[\cos(\omega_{\rm f}r_{\rm q})e^{i\omega_{\rm f}r_{\rm f}}-\cos(\omega_{\rm q}r_{\rm q})e^{-i\omega_{\rm q}r_{\rm f}}]/(2i)
$
is the coupling strength. 
Furthermore, we define the individual and correlated radiative decay rates of the qubit and the JQF as
$
\gamma_{\rm ex}^{nn} = \gamma_{\rm ex}^n\cos^2(\omega_nr_n)
$
and
$
\gamma_{\rm ex}^{\rm qf} = \gamma_{\rm ex}^{\rm fq *} = \sqrt{\gamma_{\rm ex}^{\rm q}\gamma_{\rm ex}^{\rm f}}\:[\cos(\omega_{\rm f}r_{\rm q})e^{i\omega_{\rm f}r_{\rm f}}+\cos(\omega_{\rm q}r_{\rm q})e^{-i\omega_{\rm q}r_{\rm f}}]/2
$, respectively.
The superoperator for the decay terms is defined as
${\mathcal D}(\hat{A},\hat{B})\hat{\rho}=\hat{B}\hat{\rho}\hat{A}^\dag-(\hat{A}^\dag\hat{B}\hat{\rho}+\hat{\rho}\hat{A}^\dag\hat{B})/2$.

Without loss of generality, the resonance frequencies of the qubit and JQF can be renormalized such that the Lamb shifts are included in the system Hamiltonian, i.e.\ $\hat{H}_{\rm Lamb}=0$.
In our device, the qubit is located at the open end of the semi-infinite waveguide ($r_{\rm q}=0$) and the JQF is placed at the distance $d$ apart from the qubit ($r_{\rm f}=d$).
Furthermore, the imperfections of the qubit and JQF, such as the intrinsic relaxation and the pure dephasing, are added to simulate the real device. 
Note that we assume that the control line is cooled down enough as we have checked it in the almost same configuration~\cite{Kono2018}.
As a result, the master equation of the composite system containing the qubit and JQF is described as
\begin{equation}
\label{drhodt2}
\begin{aligned}
\frac{d\hat{\rho}}{dt}= &-i\left[\hat{H}_{\rm sys}+\hat{H}_{\rm drive}+\hat{H}_{\rm eff},\:\hat{\rho}\right]
+\sum_{n,m={\rm q,f}}\gamma_{\rm ex}^{nm}\:{\mathcal D}(\hat{b}_n,\hat{b}_m)\hat{\rho}\\
&+\sum_{n={\rm q,f}}\left[\gamma_{\rm in}^n(1+n_{\rm th}^n)\:{\mathcal D}(\hat{b}_n)\hat{\rho}+\gamma_{\rm in}^nn_{\rm th}^n\:{\mathcal D}(\hat{b}_n^\dag)\hat{\rho}+\gamma_{\phi}^n\:{\mathcal D}(\hat{b}_n^\dag\hat{b}_n)\hat{\rho}\right],
\end{aligned}
\end{equation}
where $\gamma_{\rm in}^n$, $\gamma_{\phi}^n$ and $n_{\rm th}^n$ are the intrinsic relaxation rate, pure dephasing rate, and thermal quanta of the intrinsic loss channel, respectively.
The individual dissipation terms, which appear in the second line of Eq.~(\ref{drhodt2}), 
are simplified as $\:{\mathcal D}(\hat{A}) = \:{\mathcal D}(\hat{A},\hat{A})$.
Here, the system Hamiltonian in the rotating frame at drive frequency $\omega$ is given by
\begin{equation}
\label{Hsys}
\hat{H}_{\rm sys} = \sum_{n={\rm q,f}}\:\left(\omega_n - \omega\right)\hat{b}_n^\dag\hat{b}_n + \frac{\alpha_n}{2}\hat{b}_n^{\dag2}\hat{b}_n^2.
\end{equation}
The drive Hamiltonian in the rotating frame is given by 
\begin{equation}
\label{Hdrive}
\hat{H}_{\rm drive} = \sqrt{\gamma_{\rm ex}^{\rm q}\dot{n}}\left(\hat{b}_{\rm q}^\dag +\hat{b}_{\rm q} \right)+\sqrt{\gamma_{\rm ex}^{\rm f}\dot{n}}\:\cos(\omega d)\left(\hat{b}_{\rm f}^\dag+\hat{b}_{\rm f}\right).
\end{equation}
The waveguide-mediated coupling strength is given by
\begin{equation}
J = \frac{\sqrt{\gamma_{\rm ex}^{\rm q}\gamma_{\rm ex}^{\rm f}}}{2}\frac{e^{i\omega_{\rm f}d}-e^{-i\omega_{\rm q}d}}{2i}.
\end{equation}
The individual and correlated radiative decay rates to the waveguide are given by
\begin{equation}
\gamma_{\rm ex}^{\rm qq} = \gamma_{\rm ex}^{\rm q},\quad\gamma_{\rm ex}^{\rm ff} = \gamma_{\rm ex}^{\rm f}\cos^2(\omega_{\rm f}d)
\end{equation}
and 
 \begin{equation}
\gamma_{\rm ex}^{\rm qf} = \gamma_{\rm ex}^{\rm fq *} = \sqrt{\gamma_{\rm ex}^{\rm q}\gamma_{\rm ex}^{\rm f}}\:\frac{e^{i\omega_{\rm f}d}+e^{-i\omega_{\rm q}d}}{2},
\end{equation}
respectively. 
We calculate the numerical results of the dynamics of the qubit with the JQF using this master equation with the parameters listed in Table~S1.

\subsection*{Bright and dark modes}
The individual and correlated decays of the qubit and JQF can be understood as the individual decays of the bright and dark modes~\cite{Lalumiere2013}, which are defined by
\begin{equation}
\label{bBD}
\hat{b}_{\rm B/D} = \frac{\left(\gamma_{\rm ex}^{\rm B/D}-\gamma_{\rm ex}^{\rm ff}\right)\hat{b}_{\rm q}+\gamma_{\rm ex}^{\rm fq *}\:\hat{b}_{\rm f}}{\sqrt{\left(\gamma_{\rm ex}^{\rm B/D}-\gamma_{\rm ex}^{\rm ff}\right)^2+\left|\gamma_{\rm ex}^{\rm qf}\right|^2}}
\end{equation}
with decay rates
\begin{equation}
\label{gBD}
\gamma_{\rm ex}^{\rm B/D} = \frac{\gamma_{\rm ex}^{\rm qq}+\gamma_{\rm ex}^{\rm ff}}{2}\pm \sqrt{\left(\frac{\gamma_{\rm ex}^{\rm qq}-\gamma_{\rm ex}^{\rm ff}}{2}\right)^2 + \left|\gamma_{\rm ex}^{\rm qf}\right|^2}.
\end{equation}
Therefore, the radiative decay terms in Eq.~(\ref{drhodt2}) can be replaced by
\begin{equation}
\sum_{n,m={\rm q,f}}\gamma_{\rm ex}^{nm}\:{\mathcal D}(\hat{b}_n,\hat{b}_m)\hat{\rho}
\:\:\rightarrow
\sum_{\mu = {\rm B,D}}\gamma_{\rm ex}^\mu\:{\mathcal D}(\hat{b}_\mu)\hat{\rho}.
\end{equation}

\subsection*{Reflection spectrum of a single qubit}
As discussed in Sec. S5 and Sec.~S7, the qubit and JQF are first characterized individually by measuring their reflection spectra via the control line.
Therefore, we here provide a model of one of the transmon qubits (the qubit or the JQF) coupled to the semi-infinite waveguide.

By setting one of the external coupling rates of the qubit and JQF to zero in the previous model, the master equation of the single transmon qubit is obtained as
\begin{equation}
\label{drhodts}
\frac{d\hat{\rho}}{dt}= -i\left[\hat{H}_{\rm sys}+\hat{H}_{\rm drive},\:\hat{\rho}\right]
+\gamma_{\rm ex}^n\cos^2(\omega_n r_n)\:{\mathcal D}(\hat{b}_n)\hat{\rho}
+\gamma_{\rm in}(1+n_{\rm th}^n)\:{\mathcal D}(\hat{b}_n)\hat{\rho}+\gamma_{\rm in}^nn_{\rm th}^n\:{\mathcal D}(\hat{b}^\dag_n)\hat{\rho}+\gamma_{\phi}^n\:{\mathcal D}(\hat{b}^\dag_n\hat{b}_n)\hat{\rho},
\end{equation}
where
$\rho$ is the density-matrix operator of the transmon qubit,
$\hat{H}_{\rm sys} = \left(\omega_n - \omega\right)\hat{b}^\dag_n\hat{b}_n + \frac{\alpha_n}{2}\hat{b}^{\dag2}_n\hat{b}^2_n$
and
$\hat{H}_{\rm drive} = \sqrt{\gamma_{\rm ex}^n\dot{n}}\cos(\omega_n r_n)\left(\hat{b}_n^\dag +\hat{b}_n\right)$.
Note that the drive amplitude is assumed to be constant with respect to the drive frequency, i.e.\ $\cos(\omega r_n)\approx\cos(\omega_n r_n)$, since the drive frequency is close to that of the transmon qubit.

The reflection coefficient of the transmon qubit is numerically obtained by using the steady-state solution and the input-output relation.
The steady state $\hat{\rho}_{\rm ss}$ of the transmon qubit in the rotating frame of the drive frequency can be obtained by solving Eq.~(\ref{drhodts}) with $d\hat{\rho}/dt \to 0$.
Then, the expectation value of the annihilation operator of the transmon qubit in the steady state is obtained at time $t$ in the laboratory frame as
\begin{equation}
\label{bss}
\langle\hat{b}_n(t)\rangle = {\rm Tr}\left[\left(\hat{b}_n e^{-i\omega t}\right)\hat{\rho}_{\rm ss}\right] =\langle\hat{b}_n\rangle_{\rm ss}e^{-i\omega t},
\end{equation}
where $\langle\hat{b}_n\rangle_{\rm ss}={\rm Tr}\left[\hat{b}_n \hat{\rho}_{\rm ss}\right]$.
When only a single qubit is coupled to the control line, the input-output relation of Eq.~(\ref{ior}) is simplified as
\begin{equation}
\label{ior2}
\hat{a}_r =  \hat{a}_{r-t}(0) -i\sqrt{\gamma_{\rm ex}^n}\cos(\omega_n r_n)\:\hat{b}_n(t-r),
\end{equation}
where the position $r$ should be on the right-hand side of all coupling positions in the model shown in Fig.~S2(b).
Using Eq.~(\ref{bss}), the complex amplitude of the output field in the steady state can be represented as
\begin{equation}
\begin{aligned}
\label{ior3}
\langle\hat{a}_r\rangle
&= \langle\hat{a}_{r-t}(0)\rangle -i\sqrt{\gamma_{\rm ex}^n}\cos(\omega_n r_n)\:\langle\hat{b}_n(t-r)\rangle\\
&= \langle\hat{a}_{r-t}(0)\rangle -i\sqrt{\gamma_{\rm ex}^n}\cos(\omega_n r_n)\:\langle\hat{b}_n\rangle_{\rm ss}\:e^{-i\omega(t-r)}.
\end{aligned}
\end{equation}
The reflection spectrum is then obtained as
\begin{equation}
\label{S11}
{\rm S_{11}}(\omega) = \frac{\langle\hat{a}_r\rangle}{\langle\hat{a}_{r-t}(0)\rangle}
= 1-i\sqrt{\frac{\gamma_{\rm ex}^n}{\dot{n}}}\cos(\omega_n r_n)\:\langle\hat{b}_n\rangle_{\rm ss},
\end{equation}
where $\langle a_{r-t}(0) \rangle=\sqrt{\dot{n}}\:e^{-i\omega(t-r)}$ according to our earlier assumption.

\section*{S4. Working principle of JQF}
We now explain the working principle of the JQF: why the JQF suppresses the qubit radiative decay when a drive is switched off and why the JQF does not reduce the qubit Rabi frequency under a strong drive.
For simplicity, we consider a case where two-level systems are used for the qubit and JQF.
In other words, we replace the annihilation operator $\hat{b}_n$ with the lowering operator of the Pauli matrix $\hat{\sigma}_n$ in the model discussed in Sec.~S3.
After that, we consider a transmon JQF and study the dependence of the JQF anharmonicity on the Rabi frequency and Rabi decay time of the qubit.

\subsection*{Requirements for JQF}
As we discuss in the main text, we have three requirements for the JQF to act as a filter for the qubit:
\begin{quote}
 \begin{itemize}
  \item The JQF and qubit should be on resonance~($\omega_{\rm f} = \omega_{\rm q})$,
  \item The distance between the qubit and JQF should be half the qubit wavelength~($d=\lambda_{\rm q}/2$), and
  \item The JQF should be coupled to the control line much more strongly than the qubit~($\gamma_{\rm ex}^{\rm f} \gg \gamma_{\rm ex}^{\rm q})$.
 \end{itemize}
\end{quote}
The first condition~($\omega_{\rm f} = \omega_{\rm q})$ is required to maximize the cooperative effects between the qubit and JQF.
The strength of the exchange interaction and the correlated decay rate are given by $J = \frac{\sqrt{\gamma_{\rm ex}^{\rm q}\gamma_{\rm ex}^{\rm f}}}{2}\sin(\omega_{\rm q}d)$ and $\gamma_{\rm ex}^{\rm qf} = \gamma_{\rm ex}^{\rm fq} = \sqrt{\gamma_{\rm ex}^{\rm q}\gamma_{\rm ex}^{\rm f}}\cos(\omega_{\rm q}d)$, respectively. 
The second condition~($d=\lambda_{\rm q}/2$) is needed to suppress the exchange interaction, i.e.\ $J=0$, to prevent the qubit from hybridizing with the JQF, and to maximize the correlated decay, i.e.\ $\gamma_{\rm ex}^{\rm qf} = \gamma_{\rm ex}^{\rm fq} =-\sqrt{\gamma_{\rm ex}^{\rm q}\gamma_{\rm ex}^{\rm f}}$.
Using Eqs.~(\ref{bBD}) and (\ref{gBD}), the bright and dark modes are given by
\begin{equation}
\hat{\sigma}_{\rm B}=\frac{1}{\sqrt{\gamma_{\rm ex}^{\rm q}+\gamma_{\rm ex}^{\rm f}}}\left(-\sqrt{\gamma_{\rm ex}^{\rm q}}\:\hat{\sigma}_{\rm q}+\sqrt{\gamma_{\rm ex}^{\rm f}}\:\hat{\sigma}_{\rm f}\right)
\end{equation}
and 
\begin{equation}
\hat{\sigma}_{\rm D}=\frac{1}{\sqrt{\gamma_{\rm ex}^{\rm q}+\gamma_{\rm ex}^{\rm f}}}\left(\sqrt{\gamma_{\rm ex}^{\rm f}}\:\hat{\sigma}_{\rm q}+\sqrt{\gamma_{\rm ex}^{\rm q}}\:\hat{\sigma}_{\rm f}\right)
\end{equation}
with decay rates of $\gamma_{\rm ex}^{\rm B}=\gamma_{\rm ex}^{\rm q}+\gamma_{\rm ex}^{\rm f}$ and $\gamma_{\rm ex}^{\rm D}=0$.
The third condition~($\gamma_{\rm ex}^{\rm f} \gg \gamma_{\rm ex}^{\rm q}$) causes the dark mode to be close to the qubit mode, i.e. $\hat{\sigma}_{\rm D}\approx \hat{\sigma}_{\rm q}$, suppressing the radiative decay of the qubit.
On the other hand, the bright mode becomes close to the JQF mode, i.e.\ $\hat{\sigma}_{\rm B}\approx\hat{\sigma}_{\rm f}$.

\subsection*{Dynamics of qubit with JQF}
Using the bright-mode basis, the master equation~(\ref{drhodt2}) with the resonant control field ($\omega=\omega_{\rm q}$) can be rewritten as
\begin{equation}
\label{drhodtq}
\frac{d\hat{\rho}}{dt}= -i\left[\hat{H}_{\rm drive},\:\hat{\rho}\right]
+\gamma_{\rm ex}^{\rm B}\:{\mathcal D}(\hat{\sigma}_{\rm B})\hat{\rho},
\end{equation}
where $\hat{H}_{\rm drive}=-\sqrt{\gamma_{\rm ex}^{B}\dot{n}}\:(\hat{\sigma}_{\rm B}^\dag+\hat{\sigma}_{\rm B})$.
For simplicity, we neglect the imperfections of the qubit and the JQF as $\gamma_{\rm in}^n = \gamma_{\rm \phi}^n=0$.

Here, we describe the master equation with the bright- and dark modes basis: the ground state~$|{\rm gg}\rangle$,
the excited state of the bright mode~$|{\rm \widetilde{ge}}\rangle=\hat{\sigma}_{\rm B}^\dag|{\rm gg}\rangle$,
the excited state of the dark mode~$|{\rm \widetilde{eg}}\rangle=\hat{\sigma}_{\rm D}^\dag|{\rm gg}\rangle$, and
both excited states~$|{\rm ee}\rangle$.
In this representation, $|{\rm gg}\rangle$, $|{\rm ge}\rangle$, $|{\rm eg}\rangle$, and $|{\rm ee}\rangle$  are the product states of the ground and excited states of the qubit and the two-level JQF.
Using the bright- and dark-modes basis, the lowering operator of the bright mode multiplied by the coupling coefficient $\sqrt{\gamma_{\rm ex}^{\rm_B}}$, which corresponds to the transition moment of the bright mode by the control field, can be rewritten exactly as
\begin{equation}
\sqrt{\gamma_{\rm ex}^{\rm B}}\:\hat{\sigma}_{\rm B} = \sqrt{\gamma_{\rm ex}^{\rm B}}\:|\rm gg\rangle\langle{\rm \widetilde{ge}}| + \frac{\gamma_{\rm ex}^{\rm f}-\gamma_{\rm ex}^{\rm q}}{\sqrt{\gamma_{\rm ex}^{\rm B}}}\:|\rm \widetilde{eg}\rangle\langle{\rm ee}|-2\sqrt{\frac{\gamma_{\rm ex}^{\rm q}\gamma_{\rm ex}^{\rm f}}{\gamma_{\rm ex}^{\rm B}}}\:|\rm \widetilde{ge}\rangle\langle{\rm ee}|.
\end{equation}
In the limit of $\gamma_{\rm ex}^{\rm f}\gg\gamma_{\rm ex}^{\rm q}$, it can be approximated as
\begin{equation}
\begin{aligned}
\sqrt{\gamma_{\rm ex}^{\rm B}}\:\hat{\sigma}_{\rm B} &\approx \sqrt{\gamma_{\rm ex}^{\rm f}}\:|\rm gg\rangle\langle{\rm ge}| + \sqrt{\gamma_{\rm ex}^{\rm f}}\:|\rm eg\rangle\langle{\rm ee}|-2\sqrt{\gamma_{\rm ex}^{\rm q}}\:|\rm ge\rangle\langle{\rm ee}|\\
&= \sqrt{\gamma_{\rm ex}^{\rm f}}\:\hat{\sigma}_{\rm f}-\sqrt{\gamma_{\rm ex}^{\rm q}}\:(1+\hat{\sigma}_z^{\rm f})\hat{\sigma}_{\rm q}.
\end{aligned}
\end{equation}
This expression implies that the transition moment of the qubit depends on the state of the JQF, i.e.\ the qubit can be driven or decay only when the two-level JQF is in its excited state~[see Fig.~S3(a)].
Using this approximation, the master equation~(\ref{drhodtq}) can be rewritten as
\begin{equation}
\label{drhodta}
\frac{d\hat{\rho}}{dt}= -i\left[\hat{H}_{\rm drive},\:\hat{\rho}\right]
+{\mathcal D}\left(\sqrt{\gamma_{\rm ex}^{\rm f}}\:\hat{\sigma}_{\rm f}-\sqrt{\gamma_{\rm ex}^{\rm q}}\:(1+\hat{\sigma}_z^{\rm f})\hat{\sigma}_{\rm q}\right)\hat{\rho},
\end{equation}
where 
\begin{equation}
\label{approxHd}
\hat{H}_{\rm drive}=-\frac{\Omega_{\rm f}}{2}\hat{\sigma}_x^{\rm f}+\frac{\Omega_{\rm q}}{2}(1+\hat{\sigma}_z^{\rm f})\hat{\sigma}_x^{\rm q}
\end{equation}
is the approximative drive Hamiltonian where $\hat{\sigma}_x^{\rm f}$ and $\hat{\sigma}_x^{\rm q}$ are the Pauli $x$-matrix of the JQF and qubit, respectively. 
The Rabi frequencies are defined as $\Omega_{\rm f}=2\sqrt{\gamma_{\rm ex}^{\rm f}\dot{n}}$ and  $\Omega_{\rm q}=2\sqrt{\gamma_{\rm ex}^{\rm q}\dot{n}}$.

As shown in Figs.~S4(a) and (b), we numerically calculate the Rabi frequency and Rabi decay time of the qubit as a function of the control amplitude using the rigorous model (blue solid line) of Eq.~(\ref{drhodtq}) and the approximative model (orange dashed line) of Eq.~(\ref{drhodta}).
The population of the ground state of the JQF is also calculated and is shown in Fig.~S4(c).
The approximative model reproduces the results calculated with the rigorous model well.
When the control field is large enough to saturate the JQF~[see Fig.~S4(c)], the Rabi frequency with the two-level JQF approaches that without the JQF~[see Fig.~S4(a)].
On the other hand, the Rabi decay time of the qubit with the JQF is shorter than that without the JQF, even when the control field completely saturates the JQF~[see Fig.~S4(b)].

Using the approximative master equation (\ref{drhodta}), we explain the working principle of the two-level JQF.
When the JQF and qubit are driven by a control field, the drive Hamiltonian is further approximated as $\hat{H}_{\rm drive}\approx-\frac{\Omega_{\rm f}}{2}\hat{\sigma}_x^{\rm f}+\frac{\Omega_{\rm q}}{2}\hat{\sigma}_x^{\rm q}$, since the drive term $\frac{\Omega_{\rm f}}{2}\hat{\sigma}_x^{\rm f}$ suppresses the non-commutating coupling term $\frac{\Omega_{\rm q}}{2}\hat{\sigma}_z^{\rm f}\hat{\sigma}_x^{\rm q}$.
This is known as the secular approximation.
When the control field is switched off, the qubit decay immediately vanishes, since the JQF quickly decays to the ground state, resulting in $(1+\hat{\sigma}_z^{\rm f}) = 0$ in Eq.~(\ref{drhodta}).

\begin{figure}[!t]
\begin{center}
  \includegraphics[width=120mm]{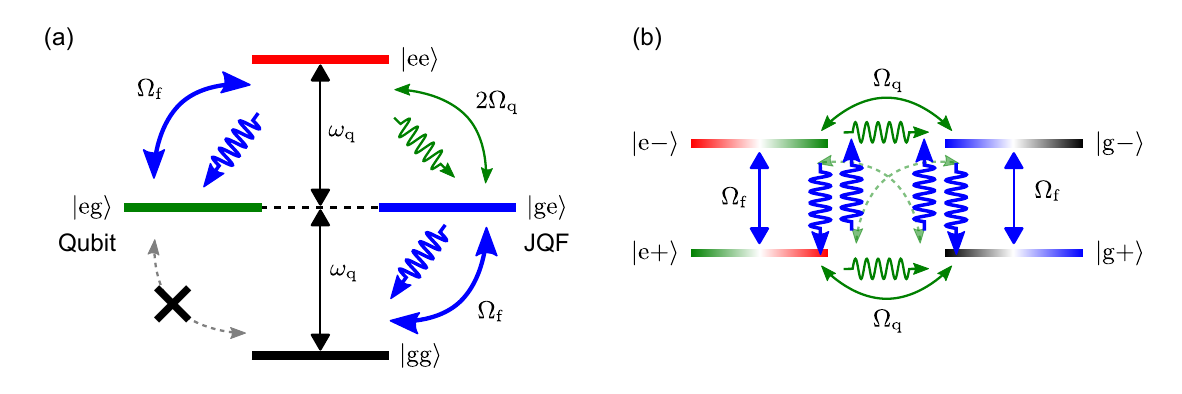} 
\caption{Schematic of energy levels.
(a)~Energy levels of the qubit and JQF in the eigenbasis of the system Hamiltonian in the laboratory frame.
The blue and green solid arrows depict the coherent interactions, while the blue and green wavy arrows show the relaxations.
The black arrows depict the qubit frequency in the laboratory frame.
(b)~Energy levels of the qubit and JQF in the basis which diagonalizes the drive Hamiltonian of the JQF in the rotating frame at drive frequency $\omega=\omega_{\rm q}$.
The blue solid arrows depict the frequency splittings due to the Rabi frequency of the JQF in the rotating frame at the drive frequency.
The green solid and dashed arrows show the resonant and off-resonant coherent interactions of the qubit, respectively.
The blue wavy arrows depict the dephasings in the Rabi oscillation of the JQF, while the green wavy arrows show the relaxations of the qubit.
}
\end{center}
\end{figure}

To understand the additional decay of the qubit Rabi oscillation caused by the two-level JQF as seen in Fig.~S4(b), we have to consider the perturbation of the coupling term $\frac{\Omega_{\rm q}}{2}\hat{\sigma}_z^{\rm f}\hat{\sigma}_x^{\rm q}$ suppressed by the secular approximation.
We explain it by using the matrix representation.

From Eq.~(\ref{approxHd}), the approximative drive Hamiltonian in the basis spanned by \{$|{\rm gg}\rangle$, $|{\rm ge}\rangle$, $|{\rm eg}\rangle$, $|{\rm ee}\rangle$\} is given by
\begin{equation}
\begin{aligned}
\hat{H}_{\rm drive}
=
\left(
\begin{array}{cccc}
0&-\frac{\Omega_{\rm f}}{2}&0&0\\
-\frac{\Omega_{\rm f}}{2}&0&0&\Omega_{\rm q}\\
0&0&0&-\frac{\Omega_{\rm f}}{2}\\
0&\Omega_{\rm q}&-\frac{\Omega_{\rm f}}{2}&0\\
\end{array}
\right).
\end{aligned}
\end{equation}
By diagonalizing the drive terms of the JQF, the Hamiltonian is represented in the basis spanned by \{$|{\rm g+}\rangle$, $|{\rm g-}\rangle$, $|{\rm e+}\rangle$, $|{\rm e-}\rangle$\} as
\begin{equation}
\label{Hdapp}
\begin{aligned}
\hat{H}_{\rm drive}
=
\left(
\begin{array}{cccc}
-\frac{\Omega_{\rm f}}{2}&0&\frac{\Omega_{\rm q}}{2}&\bcancel{-\frac{\Omega_{\rm q}}{2}}\\
0&\frac{\Omega_{\rm f}}{2}&\bcancel{-\frac{\Omega_{\rm q}}{2}}&\frac{\Omega_{\rm q}}{2}\\
\frac{\Omega_{\rm q}}{2}&\bcancel{-\frac{\Omega_{\rm q}}{2}}&-\frac{\Omega_{\rm f}}{2}&0\\
\bcancel{-\frac{\Omega_{\rm q}}{2}}&\frac{\Omega_{\rm q}}{2}&0&\frac{\Omega_{\rm f}}{2}\\
\end{array}
\right),
\end{aligned}
\end{equation}
where $|s\pm\rangle=\frac{1}{\sqrt{2}}(|s{\rm g}\rangle \pm |s{\rm e}\rangle)$ for $s={\rm g,\:e}$.
We can neglect the off-resonant coupling elements [crossed out in Eq.~(\ref{Hdapp})] in the subspaces of $|{\rm g+}\rangle$ and $|{\rm e-}\rangle$ and of $|{\rm g-}\rangle$ and $|{\rm e+}\rangle$ since the detuning $\Omega_{\rm f}$ is much larger than the coupling strength $\Omega_{\rm q}$, i.e.\ $\gamma_{\rm ex}^{\rm f}\gg\gamma_{\rm ex}^{\rm q}$, which corresponds to the secular approximation as explained above.
As schematically shown with the green solid arrows in Fig.~S3(b), the qubit Rabi oscillations between $|{\rm g+}\rangle$ and $|{\rm e+}\rangle$ and between $|{\rm g-}\rangle$ and $|{\rm e-}\rangle$ occur with the expected Rabi frequency $\Omega_{\rm q}$.

\begin{figure}[!t]
\begin{center}
  \includegraphics[width=175mm]{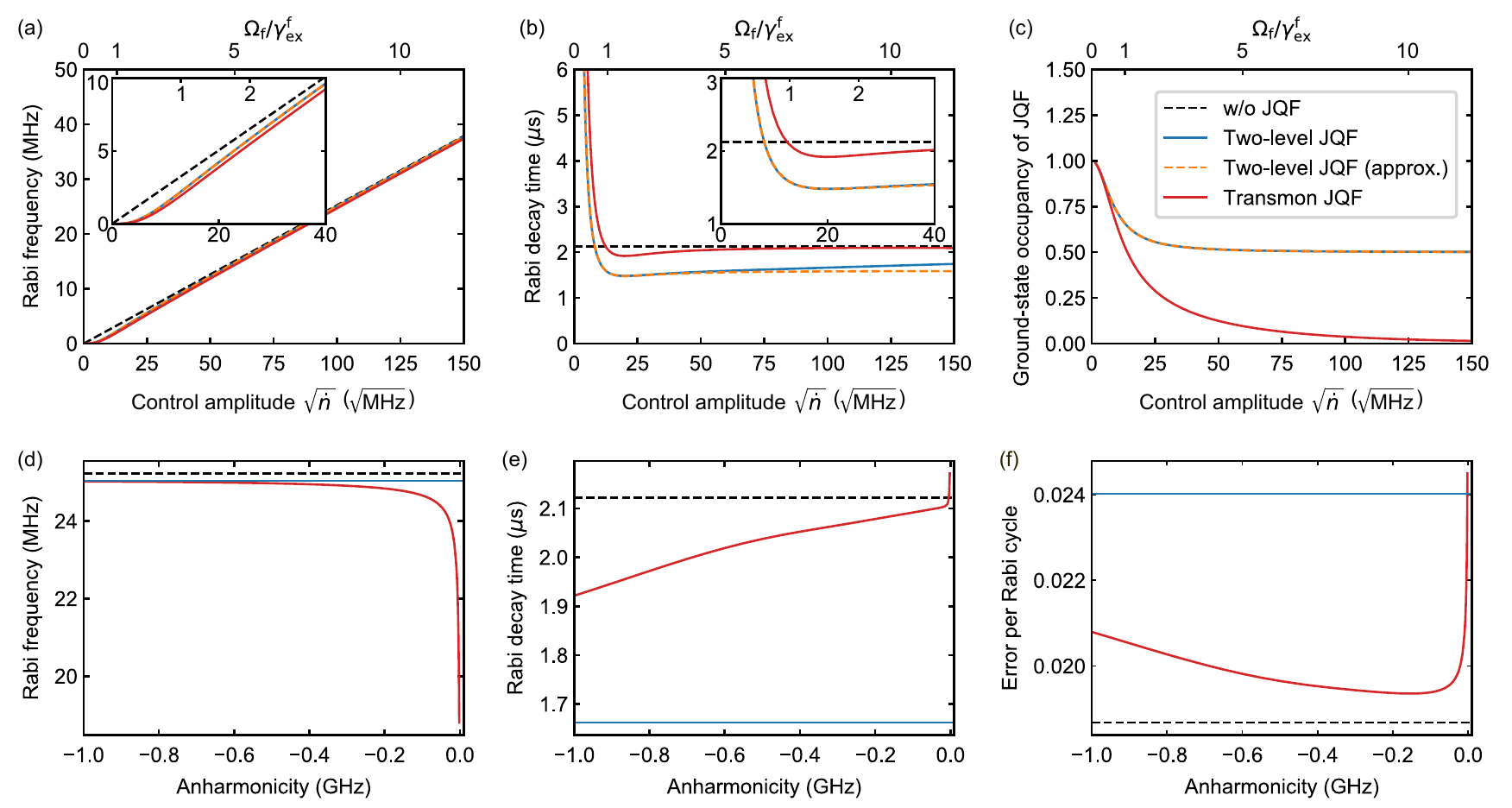} 
\caption{Numerically simulated results of qubit control in the presence of a JQF.
(a)~Rabi frequency of the qubit, (b)~Rabi decay time of the qubit, and (c)~ground-state occupancy of the JQF as a function of the control amplitude represented in the square root of the photon flux $\sqrt{\dot{n}}$.
The top axes show the corresponding Rabi frequency of the JQF $\Omega_{\rm f}=2\sqrt{\gamma_{\rm ex}^{\rm f}\dot{n}}$ normalized by its external decay rate $\gamma_{\rm ex}^{\rm f}$.
The black dashed lines are the analytical solutions with the qubit driven in the absence of the JQF. 
The blue and red solid lines are the numerically calculated results where an ideal two-level qubit and a transmon qubit with an anharmonicity of $-0.1$~GHz are used for the JQF, respectively.
The orange dashed line shows the numerically calculated results from the approximative model with a two-level JQF.
(d)~Rabi frequency, (e)~Rabi decay time, and (f)~error probability within a single Rabi cycle of the qubit as a function of the anharmonicity of the transmon-type JQF.
The system parameters used in the simulations are $\gamma_{\rm ex}^{\rm q}/2\pi = 100$~kHz, $\gamma_{\rm ex}^{\rm f}/2\pi = 100$~MHz, and  $\gamma_{\rm in}^n = \gamma_{\rm \phi}^n=0$ for $n={\rm q,\:f}$.
}
\end{center}
\end{figure}

As for the reduction of the Rabi decay time of the qubit, the additional decay channel can be explained by the same off-resonant coupling terms which we neglected using the secular approximation in the above explanation for the qubit Rabi oscillations. 
As schematically shown in Fig.~S3(b), the perturbation of the off-resonant coupling allows the qubit state to decay through the lossy JQF state~(see the green dashed arrows and the blue wavy arrows).
For example, consider the decay channel of the Rabi oscillation between $|{\rm g+}\rangle$ and $|{\rm e+}\rangle$.
First, the decay from $|{\rm e+}\rangle$ to $|{\rm g+}\rangle$ with the rate of $\gamma_{\rm ex}^{\rm q}$ corresponds to the conventional radiative decay of the qubit to the control line.
In addition, $|{\rm e+}\rangle$ ($|{\rm g+}\rangle$) decays to $|{\rm g+}\rangle$ ($|{\rm e+}\rangle$) via $|{\rm g-}\rangle$ ($|{\rm e-}\rangle$).
Here, the mixing ratio of $|{\rm g-}\rangle$ to $|{\rm e+}\rangle$ ($|{\rm e-}\rangle$ to $|{\rm g+}\rangle$) is given by $(\Omega_{\rm q}/\Omega_{\rm f})^2=\gamma_{\rm ex}^{\rm q}/\gamma_{\rm ex}^{\rm f}$.
The Rabi decay rate between $|{\rm g+}\rangle$ and $|{\rm g-}\rangle$ or between $|{\rm e-}\rangle$ and $|{\rm e+}\rangle$ is on the order of the JQF decay rate $\gamma_{\rm ex}^{\rm f}$.
Thus, these additional decay rates of the qubit are on the order of the qubit decay ($\approx(\Omega_{\rm q}/\Omega_{\rm f})^2\gamma_{\rm ex}^{\rm f}=\gamma_{\rm ex}^{\rm q}$).
As schematically shown in Fig.~S3(b), the same holds true for the Rabi oscillation between $|{\rm g-}\rangle$ and $|{\rm e-}\rangle$.

\subsection*{Transmon JQF versus two-level JQF}
In the experiments, we use a transmon qubit as the JQF.
Thus, we here consider the transmon JQF instead of the two-level JQF.
The Rabi frequency of the qubit, Rabi decay time of the qubit, and ground-state occupancy of the JQF as a function of the control amplitude are numerically calculated from the rigorous  model of Eq.~(\ref{drhodt2}), where a transmon qubit with an anharmonicity of $-0.1$~GHz is used for the JQF~[see the red lines in Figs.~S4(a)--(c)].
Note that we neglect the imperfections of the qubit and the JQF as $\gamma_{\rm in}^n = \gamma_{\rm \phi}^n=0$ for simplicity.
In the limit of large control amplitudes~($\Omega_{\rm f}/\gamma_{\rm ex}^{\rm f}\gg1$), the Rabi decay time of the qubit with the transmon JQF is increased when compared to the two-level JQF, although there is no significant difference in the qubit Rabi frequency.
As explained above, the origin of the additional qubit decay during the Rabi oscillation is the off-resonant coupling derived from the correlated decay with the lossy JQF.
In the case of a transmon JQF, the JQF can be driven to be populated in higher excited states which do not have significant correlated decay with the qubit due to the large anharmonicity.
Therefore, the Rabi decay time of the qubit with the transmon JQF can be closer to that without the JQF, as shown in Fig.~S4(b).

To further investigate the effect of the transmon-type JQF, the Rabi frequency and decay time of the qubit as a function of the anharmonicity of the transmon JQF are shown in Figs.~S4(d) and (e).
The control amplitude is set to be $\sqrt{\dot{n}}=100$~$\sqrt{\rm MHz}$.
The lower anharmonicity of the JQF results in a lower Rabi frequency of the qubit due to an effect similar to a Purcell filter.
On the other hand, since a lower anharmonicity allows for populating the higher energy levels of the transmon JQF, the Rabi decay time of the qubit approaches the same value as that without a JQF.
The error probability per single Rabi cycle is calculated as a product of the Rabi decay rate and the oscillation period~($\gamma_{\rm Rabi}^{\rm q}\times 2\pi/\Omega_{\rm q}$).
This is plotted as a function of the anharmonicity of the transmon JQF in Fig.~S4(f).
In the absence of the JQF, the lower limit of the error probability is given by $4\pi/3 \sqrt{\gamma_{\rm ex}^{\rm q}/\dot{n}}$ (black dashed line).
It is observed that the error probability is minimized when the JQF anharmonicity is close to its external coupling rate~($|\alpha_{\rm f}|\approx\gamma_{\rm ex}^{\rm f}$).

\section*{S5. JQF reflection spectrum}
The JQF is characterized by measuring the reflection spectrum via the control line.
The JQF frequency is set to be close to that of the qubit by tuning the magnetic flux applied threading the SQUID of the JQF.
Thus, we assume that $\cos(\omega_{\rm f}d)=-1$ in the model of Sec.~S3.

The amplitude and phase of the reflection coefficient as a function of the probe frequency and probe power are shown in Figs.~S5(a) and (b), while the cross-sections of the color plots are shown with different probe powers in Figs.~S5(c) and (d), respectively.
Note that the probe-frequency step is larger than the linewidth of the qubit and that the qubit resonance is not observed in this measurement.
At a smaller probe power of $-146$~dBm, the JQF spectrum is in the over-coupling regime, where the external coupling rate is much larger than the intrinsic loss rate, i.e.\ $\gamma_{\rm ex}^{\rm f}\gg\gamma_{\rm in}^{\rm f}$.
The JQF transition starts to saturate around the single-photon power level, calculated as $\hbar\omega_{\rm f}(\gamma_{\rm ex}^{\rm f}+\gamma_{\rm in}^{\rm f})^2/4\gamma_{\rm ex}^{\rm f}\approx-120$~dBm, which would populate a linear resonator with a single photon on average.
At a stronger probe power of $-100$~dBm, the JQF does not affect the reflection coefficient due to it being saturated.

We use the probe power dependence of the JQF spectrum to determine the system parameters of the JQF.
When the first excited state is thermally populated, the transition to the second excited state can be observed in our frequency range, in principle allowing us to obtain the thermal population of the JQF.
Furthermore, by fitting the reflection spectra for different powers with numerically calculated results of Eq.~(\ref{S11}), the intrinsic loss and pure dephasing rates of the JQF can be determined independently. 
We use the reflection spectra with probe powers of $-146$~dBm, $-124$~dBm and $-120$~dBm for this characterization.
Note that a phase offset and an electrical phase delay are also used as fitting parameters.
The experimental results are well reproduced by numerical calculations with the optimal fitting parameters, as shown with the black lines in Figs. S5(c) and (d).
From these fits, the system parameters of the JQF are found to be $\omega_{\rm f}/2\pi=8.0004$~GHz, $\gamma_{\rm ex}^{\rm f}/2\pi = 112$~MHz, $\gamma_{\rm in}^{\rm f}/2\pi = 3$~MHz, and $\gamma_{\rm \phi}^{\rm f}=n_{\rm th}^{\rm f}=0$.
The pure dephasing rate $\gamma_\phi^\mathrm{f}$ and the intrinsic thermal quanta $n_\mathrm{th}^\mathrm{f}$ are negligible for the JQF over-coupled to the control line.

\begin{figure}[!t]
\begin{center}
  \includegraphics[width=120mm]{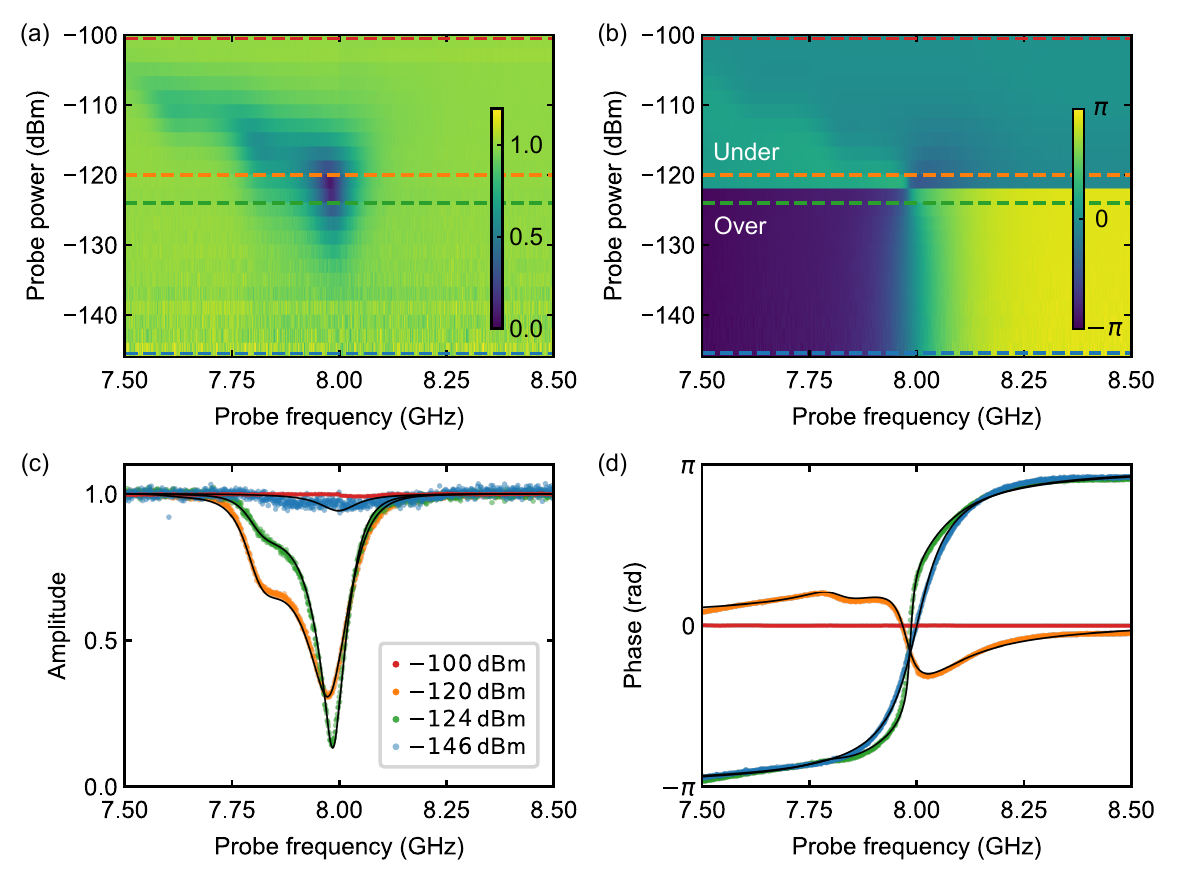} 
\caption{Reflection spectra of JQF measured via the control line.
(a)~Normalized amplitude and (b)~phase of the reflection coefficients of the JQF as functions of the probe frequency and power.
The dashed lines indicate the probe powers of the cross-sections in (c) and (d).
Note that a phase offset is added in (b) depending on the resonance being in the over- or under-coupling regimes.
(c) and (d) Cross sections of (a) and (b) at different probe powers. 
The dots are the experimental results while the black lines are the theoretical fits for $-146$~dBm, $-124$~dBm and $-120$~dBm.
}
\end{center}
\end{figure}

\begin{figure}[!t]
\begin{center}
  \includegraphics[width=160mm]{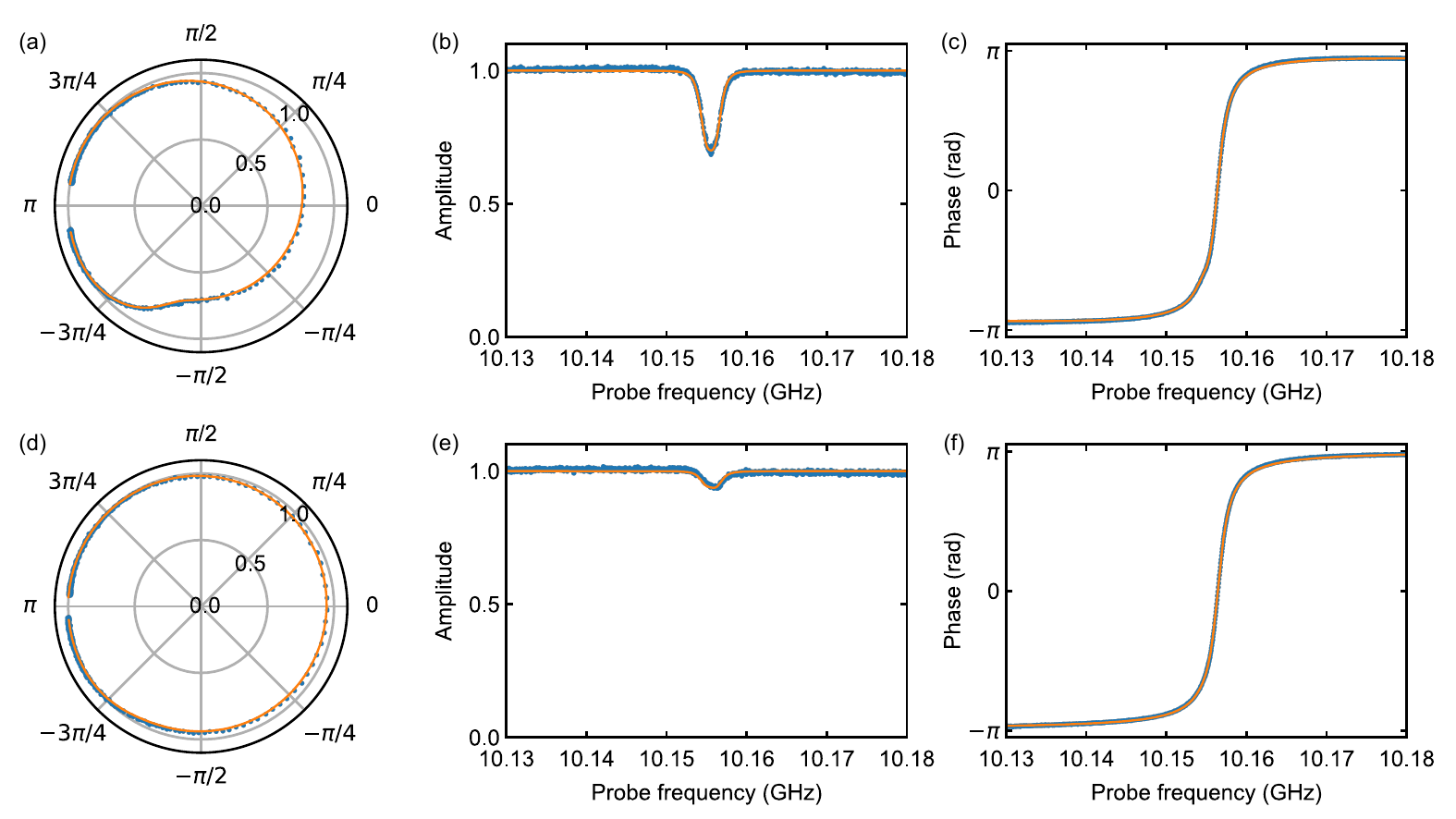} 
\caption{Reflection spectra of the readout resonator.
The dots and solid lines are the experimental results and the theoretical fits, respectively.
(a)~Complex amplitude, (b)~normalized amplitude and (c)~phase of the reflection spectrum of the readout resonator when the JQF is resonant with the qubit.
(d)-(f)~Same plots for the JQF out of resonance from the qubit.
}
\end{center}
\end{figure}

\section*{S6. Readout-resonator reflection spectrum}
To characterize the thermal population of the qubit, we observe the dispersive frequency shift of the readout resonator depending on the qubit state. 

The reflection spectrum of a resonator is given by
\begin{equation}
S_{11}^{\:\rm r}(\omega,\omega_{\rm r},\kappa_{\rm ex},\kappa_{\rm in}) =  -\frac{\frac{\kappa_{\rm ex}-\kappa_{\rm in}}{2}+i(\omega-\omega_{\rm r})}{\frac{\kappa_{\rm ex}+\kappa_{\rm in}}{2}-i(\omega-\omega_{\rm r})},
\end{equation}
where $\omega$ is the probe frequency, $\omega_{\rm r}$ is the resonance frequency, $\kappa_{\rm ex}$ is the external coupling rate, and $\kappa_{\rm in}$ is the intrinsic loss rate~\cite{Clerk2010}.

In our setup, the readout resonator is coupled to the transmon qubit dispersively. Thus, the resonance frequency of the readout resonator is shifted by the dispersive shift $\chi$ depending on the qubit state: the resonance frequency is $\omega_{\rm r}+\chi$ ($\omega_{\rm r}-\chi$) when the qubit is in the ground (first excited) state~\cite{Gambetta2006}.

The measurement time of the reflection signal is much longer than the time scale of the qubit dynamics in thermal equilibrium, i.e. the qubit state is hopping between the ground and first excited states during the measurement. The ratio of the dwell times of each state corresponds to the qubit thermal population.
We assume that the second excited state of the qubit does not get populated.
The reflection spectra of the readout resonator with the qubit in the ground and first excited states are then classically mixed as
\begin{equation}
\label{S11ge}
S_{11}^{\rm ge}(\omega,\omega_{\rm r},\kappa_{\rm ex},\kappa_{\rm in}, \chi, p_{\rm th}^{\rm q}) = (1-p_{\rm th}^{\rm q})\:S_{11}^{\:\rm r} (\omega, \omega_{\rm r}+\chi,\kappa_{\rm ex},\kappa_{\rm in}) + p_{\rm th}^{\rm q}\:S_{11}^{\:\rm r}(\omega,\omega_{\rm r}-\chi,\kappa_{\rm ex},\kappa_{\rm in}),
\end{equation}
where $p_{\rm th}^{\rm q}$ is the qubit thermal population in the first excited state.
We use the analytical solution of the readout resonator to fit the experimental results.
In this fitting, the intrinsic loss rate of the resonator and the thermal population of the qubit can be determined independently when the coupled system is in the strong dispersive regime, i.e.\ $\kappa_{\rm ex}+\kappa_{\rm in}\leq2\chi$.
Note that a phase offset and an electrical phase delay are also used as additional fitting parameters.

As shown in Figs.~S6(a)--(c), the complex amplitude, normalized amplitude, and phase of the reflection spectrum of the readout resonator are measured when the JQF is on resonance with the qubit, respectively.
Note that the probe power is set to be $-130$~dBm, comparable with the single-photon power level for the readout resonator, defined as $\hbar\omega_{\rm r}(\kappa_{\rm ex}+\kappa_{\rm in})^2/(4\kappa_{\rm ex})\approx-136$~dBm.
The reflection spectrum of the readout resonator is fitted well with Eq.~(\ref{S11ge}).
Since we assume that the control line is well cooled down at the qubit frequency, the thermal population of the qubit is maximized due to the hotter intrinsic bath when the JQF is on resonance with the qubit and decouples the qubit from the control line.
Therefore, the ratio of the reflection spectrum with the qubit in the excited state is larger, which enables us to determine the resonator frequency and dispersive shift more precisely as $\omega_{\rm r}/2\pi = 10.1564$~GHz and $2\chi/2\pi= 1.870$~MHz.

As shown in Figs.~S6(d)--(f), the reflection spectrum of the readout resonator is measured when the JQF is out of resonance from the qubit, enabling us to characterize the thermal population of the qubit in the absence of the JQF. 
By fitting the reflection spectrum with the determined parameters of $\omega_{\rm r}$ and $\chi$, the external coupling and intrinsic loss rates of the resonator and the thermal population of the qubit are determined  as $\kappa_{\rm ex}/2\pi = 2.152$~MHz, $\kappa_{\rm in}/2\pi = 0.015$~MHz, and $p_{\rm th}^{\rm q}=0.028$.
As shown in Fig.~S9(c), the thermal population of the qubit with different JQF-qubit detunings is determined in the same way.
We will use the thermal population of the qubit to determine the external coupling rate of the qubit in the next section.

\section*{S7. Qubit reflection spectrum}
To characterize the external coupling rate of the qubit, the reflection spectrum of the qubit is measured via the control line. The external coupling rate of the qubit in the absence of the JQF is used for the calibration of the control power~(see Sec~S8).

Since the qubit linewidth is much smaller than its anharmonicity, under the assumption of no thermal excitation of its second excited state, the transmon qubit can be well approximated as a two-level system, i.e.\ the truncation number of the annihilation operator is set to 2 in the model discussed in Sec.~S3.
Therefore, the off-diagonal element of the steady state of the master equation~(\ref{drhodts}) is analytically solved as
\begin{equation}
\langle 0|\hat{\rho}_{ss} |1\rangle = i\frac{\sqrt{\gamma_{\rm ex}^{\rm q}\dot{n}}}{(2n_{\rm eff}^{\rm q}+1)\gamma_2^{\rm q}}\:\frac{1+i\left(\frac{\omega-\omega_{\rm q}}{\gamma_2^{\rm q}}\right)}{1+\left(\frac{\omega-\omega_{\rm q}}{\gamma_2^{\rm q}}\right)^2+\frac{4\gamma_{\rm ex}^{\rm q}\dot{n}}{\gamma_1^{\rm q}\gamma_2^{\rm q}}
},
\end{equation}
where $n_{\rm eff}^{\rm q}=\gamma_{\rm in}^{\rm q}n_{\rm th}^{\rm q}/(\gamma_{\rm ex}^{\rm q}+\gamma_{\rm in}^{\rm q})$ is the effective thermal quanta,
$\gamma_1^{\rm q}= \gamma_{\rm ex}^{\rm q}+(2n_{\rm th}^{\rm q}+1)\gamma_{\rm in}^{\rm q}$ is the total relaxation rate, and 
$\gamma_2^{\rm q} = \gamma_1^{\rm q}/2+\gamma_{\rm \phi}^{\rm q}$ is the total dephasing rate~\cite{Mirhosseini2019}.
Using Eq.~(\ref{S11}), the reflection spectrum of the qubit is given by
\begin{equation}
S_{11}^{\:\rm q} = 1- \frac{\gamma_{\rm ex}^{\rm q}}{(2n_{\rm eff}^{\rm q}+1)\gamma_2^{\rm q}}\:\frac{1+i\left(\frac{\omega-\omega_{\rm q}}{\gamma_2^{\rm q}}\right)}{1+\left(\frac{\omega-\omega_{\rm q}}{\gamma_2^{\rm q}}\right)^2+\frac{4\gamma_{\rm ex}^{\rm q}\dot{n}}{\gamma_1^{\rm q}\gamma_2^{\rm q}}},
\end{equation}
where $\langle\hat{b}_{\rm q}\rangle_{\rm ss}=\langle 0|\hat{\rho}_{\rm ss} |1\rangle$ is used.
In the limit of weak probe power, i.e.\ $\dot{n}\ll\gamma_1^{\rm q}\gamma_2^{\rm q}/(4\gamma_{\rm ex}^{\rm q})$, the reflection spectrum can be approximated as
\begin{equation}
\label{S11q}
S_{11}^{\:\rm q}(\omega,\omega_{\rm q},\gamma_{\rm eff}^{\rm q},\gamma_2^{\rm q}) = 1-\frac{\gamma_{\rm eff}^{\rm q}}{\gamma_2^{\rm q}-i(\omega-\omega_{\rm q})},
\end{equation}
where an effective rate $\gamma_{\rm eff}^{\rm q}= \gamma_{\rm ex}^{\rm q}/(2n_{\rm eff}^{\rm q}+1)$ is defined as an independent fitting parameter.
Note that a phase offset and an electrical phase delay are also used as fitting parameters.
Furthermore, from the experimentally-obtainable qubit thermal population $p_{\rm th}^{\rm q}$, the effective thermal quanta $n_{\rm eff}^{\rm q}$ is obtained as
\begin{equation}
n_{\rm eff}^{\rm q} = \frac{p_{\rm th}^{\rm q}}{1-2p_{\rm th}^{\rm q}}.
\end{equation}
As a result, the external coupling rate of the qubit is determined to be
\begin{equation}
\label{gam_exq}
\gamma_{\rm ex}^{\rm q} = (2n_{\rm eff}^{\rm q}+1) \gamma_{\rm eff}^{\rm q}.
\end{equation}

\begin{figure}[!t]
\begin{center}
  \includegraphics[width=170mm]{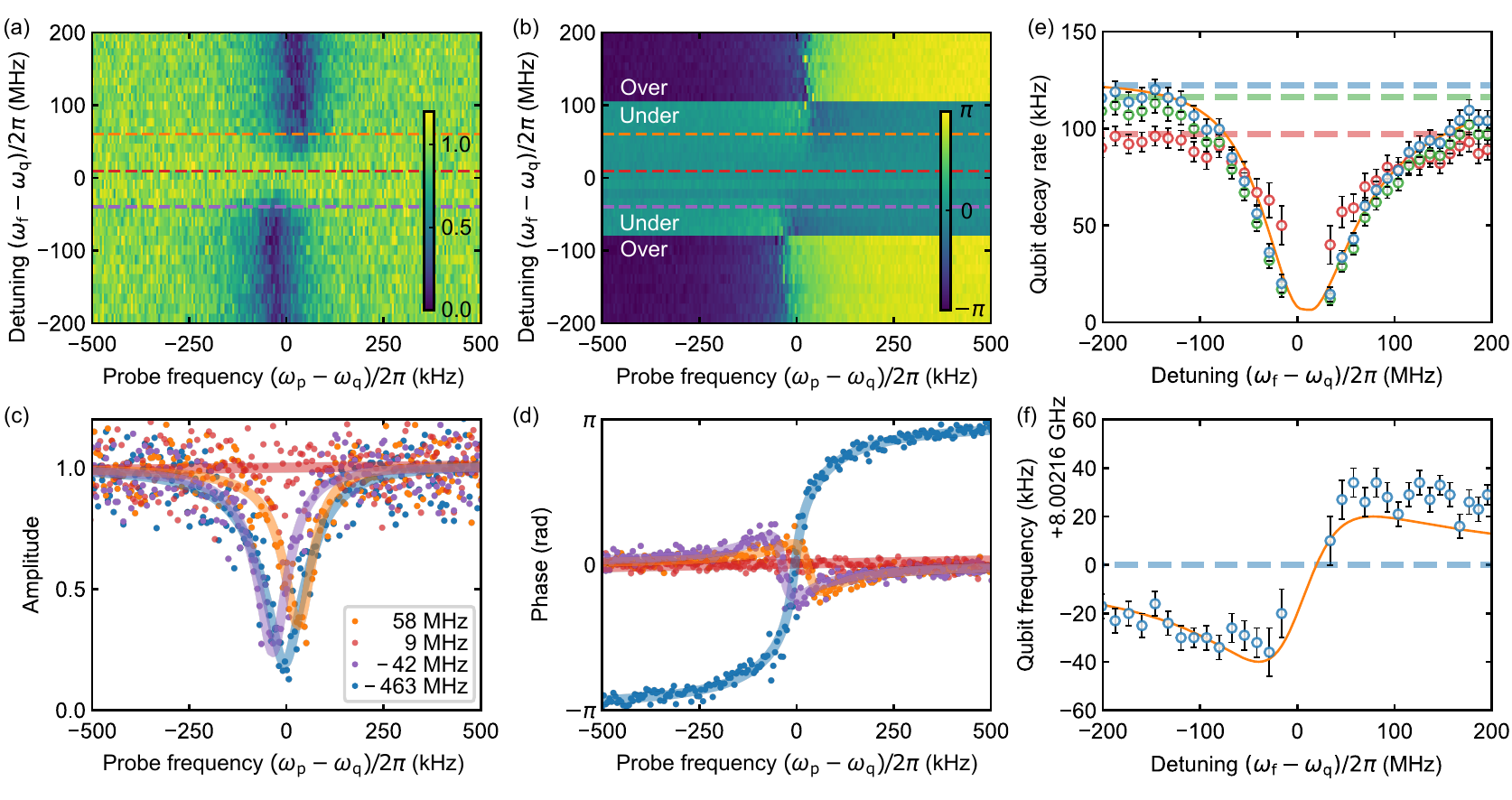} 
\caption{Reflection spectra of the qubit measured via control line.
(a)~Amplitude and (b)~phase of the reflection coefficients as functions of the probe-qubit and JQF-qubit detunings.
The dashed lines indicate the detunings of the cross-sections in (c) and (d).
Note that a phase offset is added in (b) depending on the resonance being in the over- or under-coupling regimes.
(c) and (d) Cross sections of (a) and (b) at different JQF-qubit detunings. 
The dots and solid lines are the experimental results and the theoretical fits, respectively.
(e)~Qubit decay rates as a function of the detuning.
The green, red, and blue circles are the fitting results of the effective rate $\gamma_{\rm eff}^{\rm q}$, total dephasing rate $\gamma_2^{\rm q}$, and corrected external coupling rate $\gamma_{\rm ex}^{\rm q}$, respectively.
The dashed lines depict the experimental results in the absence of the JQF.
(f)~Qubit frequency as a function of the JQF-qubit detuning.
The dashed line is the qubit frequency experimentally obtained in the absence of the JQF.
The orange solid lines in (e) and (f) are the external coupling rate and resonance frequency of the qubit which are numerically obtained by the master equation Eq.~(\ref{drhodt2}), respectively.
}
\end{center}
\end{figure}

The amplitude and phase of the reflection coefficient as a function of the probe frequency and the JQF-qubit detuning are shown in Figs.~S7(a) and (b), respectively.
To obtain the linear response of the qubit, we set the probe power to $-166$~dBm in these measurements.
The qubit spectrum with the far detuning from the JQF (in the absence of the JQF) is found to be in the over-coupling regime to the control line.
In contrast, when the JQF is on resonance with the qubit, the signature of the qubit transition disappears, indicating that the qubit is decoupled from the control line.

As shown in Figs.~S7(c) and (d), the qubit reflection spectra with the different JQF-qubit detunings are fitted well with the theoretical model of Eq.~(\ref{S11q}).
The fitting results as a function of the detuning are plotted in Figs.~S7(e) and (f).
The external coupling rate of the qubit, which is determined from Eq.~(\ref{gam_exq}), is also shown in Figs.~S7(e).
Note that we use the thermal population of the qubit for the various detuning, which is obtained from the numerical simulation, as shown in Fig.~S9(c).
The full frequency bandwidth at half maximum of the suppression spectrum is found to be about $130$~MHz, which roughly coincides with the external coupling rate of the JQF.
The asymmetry of the suppression spectrum is explained by the non-ideal distance between the qubit and the JQF, i.e.\ $d=0.526\lambda_{\rm q}$~(see Sec.~S9).

The numerically calculated external coupling rate and resonance frequency of the qubit as a function of the detuning are shown with the orange solid lines in Figs.~S5(e) and (f), respectively.
Here, the numerical results reproduce the experimental results without any fitting parameters.
The external coupling rate of the qubit in the presence of the JQF is numerically obtained by calculating the relaxation rate of the qubit with the master equation Eq.~(\ref{drhodt2}) where the intrinsic loss rate is set to $\gamma_{\rm in}^{\rm q}=0$.
The qubit frequency is numerically obtained by fitting the simulation results of the Ramsey sequence with a damped sinusoidal curve.

The bare external coupling rate of the qubit is used for the calibration of the qubit control power as discussed in Sec.~S8.
When the JQF is far detuned from the qubit, the effective rate is determined as $\gamma_{\rm eff}^{\rm q}/2\pi = 116$~kHz. 
Using $p_{\rm th}^{\rm q}=0.028$ as found in Sec.~S6, the effective thermal quanta is determined as $n_{\rm eff}^{\rm q}=0.029$. 
Then, the qubit external coupling rate in the absence of the JQF is determined as $\gamma_{\rm ex}^{\rm q}/2\pi = 123$~kHz.

\begin{figure}[!t]
\begin{center}
  \includegraphics[width=70mm]{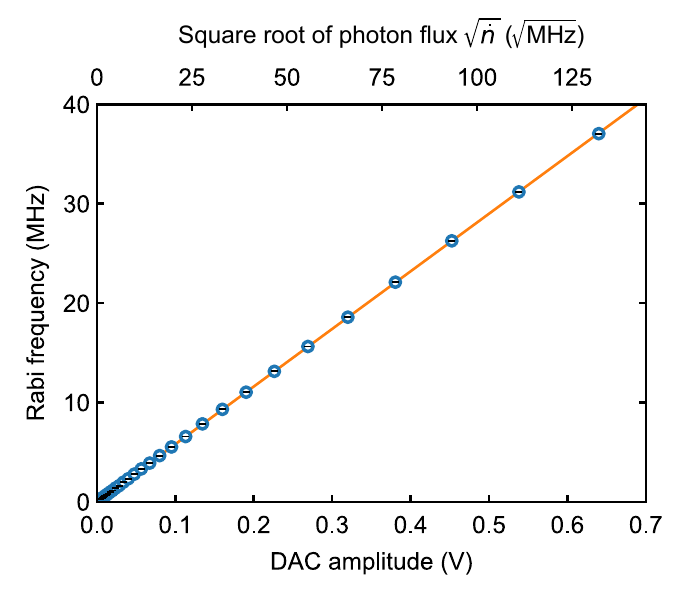} 
\caption{Qubit Rabi frequency as a function of the control amplitude without the JQF.
The dots and the solid line are the experimental results and the linear fit, respectively.
The bottom and top scales describe the DAC amplitude and the calibrated control amplitude represented in the square root of the photon flux $\sqrt{\dot{n}}$, respectively.
}
\end{center}
\end{figure}

\begin{figure}[!t]
\begin{center}
  \includegraphics[width=175mm]{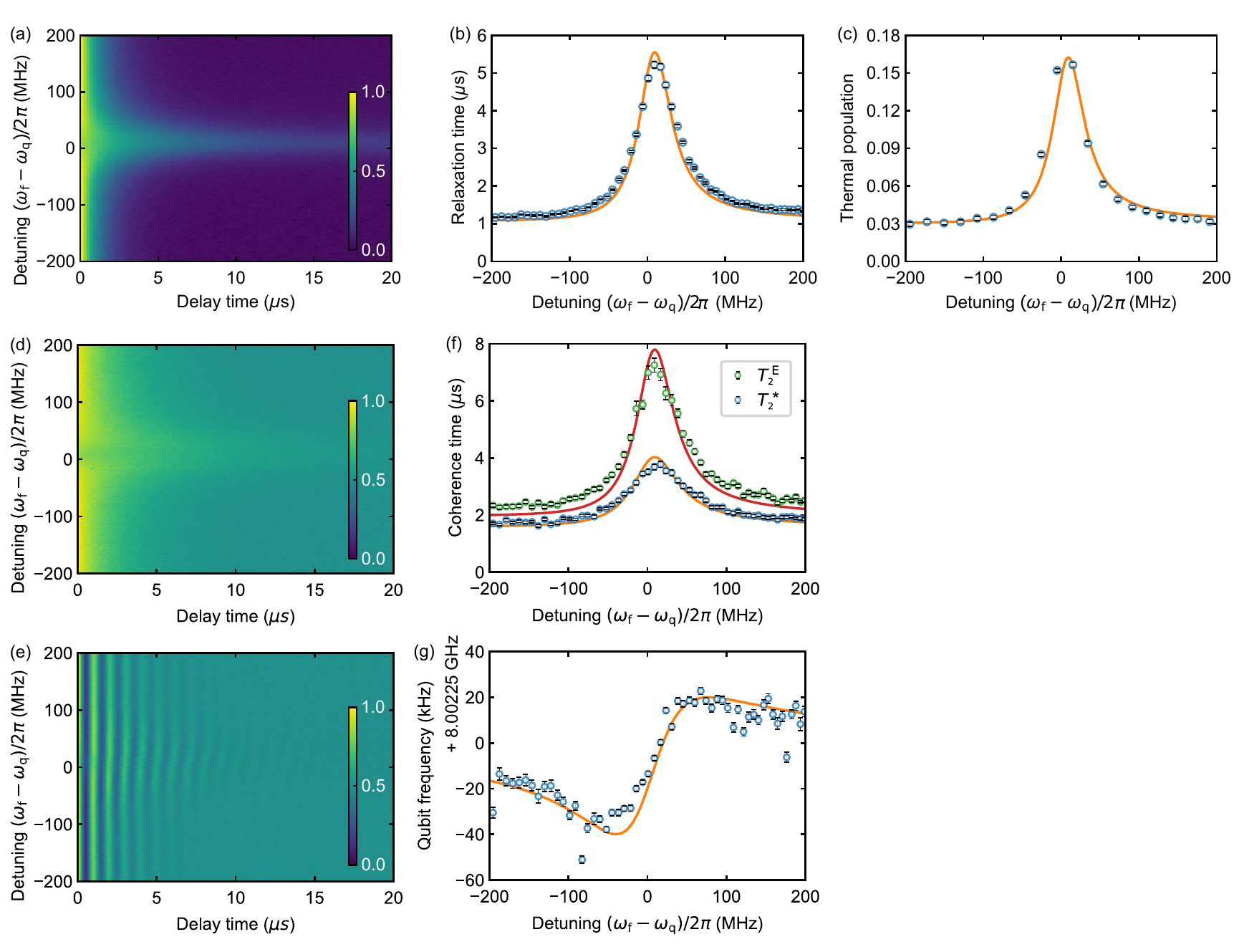} 
\caption{Qubit relaxation, coherence and thermal population.
(a)~Qubit relaxation as a function of the JQF-qubit detuning.
The color scales represent the qubit populations in the excited state.
(b)~Relaxation time and (c)~thermal population of the qubit as a function of the detuning.
The dots and solid lines are the experimental results and the theoretical fits, respectively.
(d) and (e)~Coherence of the qubit with the Hahn-echo and Ramsey sequences as a function of the detuning, respectively.
(f)~Coherence times and (g)~resonance frequency of the qubit as functions of the detuning.
The circles and lines are the experimental results and the numerical calculations, respectively.
}
\end{center}
\end{figure}

\section*{S8. Qubit Rabi oscillation}
To calibrate the photon flux of the control microwave field, the qubit Rabi frequency is measured in the absence of the JQF.
Rabi oscillations are observed by applying the duration-varying control pulse followed by the readout pulse with the JQF frequency set to be far detuned from the qubit frequency.
By fitting the experimental results with a damped sinusoidal curve, the Rabi frequency is obtained as a function of the amplitude of the control signal generated by the digital-analog converter~(DAC), as shown in Fig.~S8.
Usually, it is hard to precisely determine the ratio of the square root of the control photon flux to the DAC amplitude from the experimental setup.
Here, this ratio is determined by comparing the slope of the Rabi frequency as a function of the DAC amplitude with $2\sqrt{\gamma_{\rm ex}^{\rm q}}$, since the Rabi frequency is given by $\Omega_{\rm q}=2\sqrt{\gamma_{\rm ex}^{\rm q}\dot{n}}$.
Using the external coupling rate of the qubit in the absence of the JQF, which is determined in Sec.~S7, the square root of the photon flux for a given DAC amplitude is calibrated as shown with the upper scale in Fig.~S8.

\section*{S9. Qubit decay times with JQF}
Here, we characterize the intrinsic loss rate $\gamma_{\rm in}^{\rm q}$, pure dephasing rate $\gamma_{\rm \phi}^{\rm q}$ and intrinsic thermal quanta $n_{\rm th}^{\rm q}$ of the qubit by measuring its relaxation and coherence as a function of the JQF-qubit detuning.

As shown in Fig.~S9(a), the qubit relaxation is measured as a function of the JQF-qubit detuning.
By fitting the experimental results with an exponential curve, the relaxation time of the qubit is determined as shown in Fig.~S9(b).
The thermal population of the qubit, which is obtained from the reflection spectrum of the readout resonator as discussed in Sec.~S6, is shown as a function of the detuning in Fig.~S9(c).
Since not only the intrinsic loss but also the thermal population contributes to the qubit relaxation time, the intrinsic loss rate $\gamma_{\rm in}^{\rm q}$ and the intrinsic thermal quanta $n_{\rm th}^{\rm q}$ are determined simultaneously by fitting the relaxation time and thermal population of the qubit with numerical calculations based on the master equation (\ref{drhodt2}).
The thermal population is numerically obtained from the steady state of the master equation without a drive field.
In addition, since the asymmetric shapes of the peaks in Figs.~S9(b) and (c) is explained by the mismatch between the ideal distance of the half-wavelength and the actual distance, the distance $d$ is considered as a fitting parameter.
The experimental results are reproduced well by numerical simulations with optimal fitting parameters, as shown in Figs.~S9(b) and (c).
The system parameters are determined as $\gamma_{\rm in}^{\rm q}/2\pi=16$~kHz, $n_{\rm th}^{\rm q} = 0.29$, and $d=0.526\lambda_{\rm q}$, where $\lambda_{\rm q}$ is the wavelength at the qubit frequency.

Furthermore, we measure the qubit coherence as a function of the JQF-qubit detuning.
Fig.~S9(f) shows the qubit coherence times which are determined by fitting the experimental results obtained by the Hahn-echo and Ramsey sequences~[shown in Figs.~S9(d) and (e)] with exponential and damped sinusoidal curves, respectively.
In Fig.~S9(g), we show the qubit frequency which is determined from fitting the data of the Ramsey sequence with a finite detuning between the qubit frequency and the control frequency~[shown in Figs.~S9(e)].
The coherence times of the qubit as a function of the detuning are fitted with the numerical calculation results obtained by the master equation~(\ref{drhodt2}), where the pure dephasing rate of the qubit is considered as a fitting parameter.
In the numerical simulation, the qubit is initialized in the coherent superposition state and the qubit population in the $x$ basis is calculated after the varying delay time without a drive field. 
Note that the rotating frame of the numerical calculation is shifted from the bare qubit frequency by a finite detuning in order to easily distinguish the dephasing from the frequency shift of the qubit.
The experimental results agree well with the numerical calculation ones, enabling us to determine the pure dephasing rate to be $\gamma_{\rm \phi}^{\rm q}/2\pi=6$~kHz for the Hahn-echo sequence and $\gamma_{\rm \phi}^{\rm q}/2\pi=25$~kHz for the Ramsey sequence, respectively.
We use the pure dephasing rate from the Hahn-echo sequence for the simulation of the Rabi oscillation and the calculation of the coherence limit of the average gate error of the Clifford gates.
The qubit frequency as a function of the detuning is also reproduced well by the numerical calculation, as shown in Fig.~S9(g).

\clearpage
\end{widetext}

\end{document}